\documentclass[lettersize,journal]{IEEEtran}
\usepackage{amsmath,amssymb,amsfonts}
\usepackage{algorithmic}
\usepackage{algorithm}
\usepackage{array}
\usepackage{textcomp}
\usepackage{stfloats}
\usepackage{url}
\usepackage{verbatim}
\usepackage{graphicx, caption, subcaption}
\usepackage{cite}
\usepackage{xcolor}
\usepackage{color,soul}
\usepackage{pifont}
\usepackage{multirow}
\usepackage{multicol}

\begin{document}

\title{Towards Scalable Multi-Chip Wireless Networks with Near-Field Time Reversal
\thanks{\textsuperscript{*} A. Bandara and F. Rodríguez-Galán contributed equally to this work.
\newline Authors acknowledge support from the European Union’s Horizon Europe research and innovation program under grant agreement 101042080 (WINC) as well as the European Innovation Council (EIC) PATHFINDER scheme, grant agreement No 101099697 (QUADRATURE). }
}

\author{Ama Bandara*, \and F\'atima Rodr\'iguez-Gal\'an*, \and Pau Talarn, \and Elana Pereira de Santana, \and Evgenii Vinogradov,
\and Peter Haring Bol\'ivar, \and Eduard Alarc\'on, \and Sergi Abadal
\thanks{A.Bandara, F. Rodríguez-Galán, P. Talarn, Evgenii Vinogradov, E. Alarcón, S.  Abadal are with Universitàt Politècnica de Catalunya, Spain. E. Pereira de Santana and P. Haring Bolívar are with University of Siegen, Germany}
}

\maketitle

\begin{abstract}
The concept of Wireless Network-on-Chip (WNoC) has emerged as a potential solution to address the escalating communication demands of modern computing systems due to its low-latency, versatility, and reconfigurability. However, for WNoC to fulfill its potential, it is essential to establish multiple high-speed wireless links across chips. Unfortunately, the compact and enclosed nature of computing packages introduces significant challenges in the form of Co-Channel Interference (CCI) and Inter-Symbol Interference (ISI), which not only hinder the deployment of multiple spatial channels, but also severely restrict the symbol rate of each individual channel. In this paper, we posit that Time Reversal (TR) could be effective in addressing both impairments in this static scenario, thanks to its spatiotemporal focusing capabilities even in the near-field. Through comprehensive full-wave simulations and bit error rate analysis in multiple chip layouts with multiple frequency bands, we provide evidence that TR can increase the symbol rate by an order of magnitude, enabling the deployment of multiple concurrent links and achieving aggregate speeds exceeding 100 Gb/s. Finally, we evaluate the impact of reducing the sampling rate of the TR filter on the achievable speeds, paving the way to practical TR-based wireless communications at the chip scale. 
\end{abstract}


\section{Introduction}
\label{sec:intro}

The end of Moore's law has compelled computer architects to explore methods for improving the performance and efficiency of microprocessors beyond mere transistor scaling. One approach involves increasing the number of independent processor cores within a chip, resulting in what are known as many-core processors \cite{Nychis2012}. Alternatively, other strategies encompass the deployment of specialized hardware accelerators and the integration of multiple such specialized chips within a computing package \cite{zimmer20200}. Given the parallel nature of these architectures and their requirement for internal data synchronization and sharing, effective communication within and between chips emerges as a pivotal factor determining the performance and efficiency of next-generation computing systems.

Currently, numerous processor families in both academic and industrial settings adopt the Network-on-Chip (NoC) as the prevailing solution for on-chip communications \cite{Nychis2012}. A NoC functions as a packet-switched network comprising on-chip routers and links co-integrated into the architecture. Its widespread use is attributed to its simplicity and high performance, particularly in moderately sized processors. However, as architectures scale towards larger sizes and extend beyond the confines of a single chip, where wired links experience significant slowdowns \cite{zimmer20200}, NoC encounters notable challenges related to latency and energy consumption. This becomes especially problematic during collective communications, such as one-to-all and all-to-one patterns, which can lead to severe congestion within the NoC.

In this context, Wireless Network-on-Chip (WNoC) is an emerging paradigm with significant promise to enhance communications within computing packages. WNoC are possible because of the advances in antennas and transceivers in the millimeter-wave (mmWave) and terahertz (THz) bands that have made possible their integration inside computing cores \cite{cheema2013last}. They are expected to mitigate the scalability and flexibility issues that come with the increment of the number of processors, their integration on multiple interconnected chips, and the use of wired NoC communication approaches \cite{Nychis2012, Todri-Sanial2017}. The goal is to augment the wired backbone with wireless links within the package \cite{timoneda2020engineer}.

Due to the use of electromagnetic (EM) propagation nearly at the speed of light, WNoC provides broadcast and multicast opportunities with reduced latency and is expected to offer reconfigurable links able to operate at speeds of tens of Gb/s or more. For the WNoC paradigm to be competitive, though, it must deliver a reliability commensurate to that of their wired counterpart. In particular, multiple works \cite{Sujay2012,Shamim2017,abadal2018medium} point towards a latency of 1--10 nanoseconds, a throughput in the range of 10--100 Gb/s or higher, and a Bit Error Rate (BER) of 10\textsuperscript{-12}--10\textsuperscript{-15} as potential targets for WNoC. Complying with these values while keeping low the area/power consumption calls for affordable solutions, low order modulations, and simple physical layer elements for implementation \cite{Pano2020a, Shamim2017,Matolak2013CHANNEL}.

However, several challenges arise when including wireless communications in a package environment while trying to meet the target performance requirements. Some examples include the antenna placement in the highly integrated environment, design of transceivers with efficient coding schemes and modulations, interference management, and medium access control \cite{timoneda2020engineer, jornet2014femtosecond, mansoor2015reconfigurable}. 
In particular, in this paper we focus on the problems of Inter-Symbol Interference (ISI) and Co-Channel Interference (CCI). In the on-chip scenario, the antennas radiate in a highly reverberant environment and are impaired by near-field effects. As a result, on the one hand, signals are subject to multiple reflections before reaching the receiver, leading to large delay spreads dominated by the strong multipath components. This leads to severe ISI, which degrades the performance of high-rate transmissions especially in low-order modulations. On the other hand, reverberation causes the radiated energy to spread across the entire space. This leads to strong CCI with non-intended antennas and prevents the possibility of having multiple spatial channels concurrently. Other techniques such as Orthogonal Frequency Division Multiple Access (OFDMA) and beamforming could address ISI and CCI, but are not advisable in the chip-scale communications scenario due to the required complex and power-hungry transceivers, as well as the need for multiple antennas consuming a large area, respectively \cite{yang2010ofdma, Aslan2021}.   

Time Reversal (TR) \cite{lerosey2004time, lerosey2007focusing, Han2012, alexandropoulos2022time} is a technique that uses a time-reversed version of the channel's impulse response (CIR) to create a matched filter that focuses energy in space and time in a targeted destination. Due to the spatiotemporal focusing, TR can therefore considerably alleviate both ISI and CCI, as shown in multiple works at multiple scales and including in the near-field \cite{Chabalko2016, lerosey2007focusing}. 

In traditional large-scale wireless systems, this approach is expensive to implement because the channel, being dynamic, would have to be estimated continuously. In the chip scenario, however, the node position is fixed and every element in the system is known. This works to our advantage because regardless of the channel complexity, this will be almost fixed and static. One estimation of the CIR of the wireless path between terminals of interest will remain constant and can be used in every transmission. Moreover, the near-field effects naturally arising in WNoC due to the small scale of the chip environment are assumed to be negligible, since TR can be applied to channels with near-field effects as long as their consequences are captured in the CIR \cite{Chabalko2016, lerosey2007focusing}. 

\begin{figure}[t]
\centering
\includegraphics[width=0.5\textwidth]{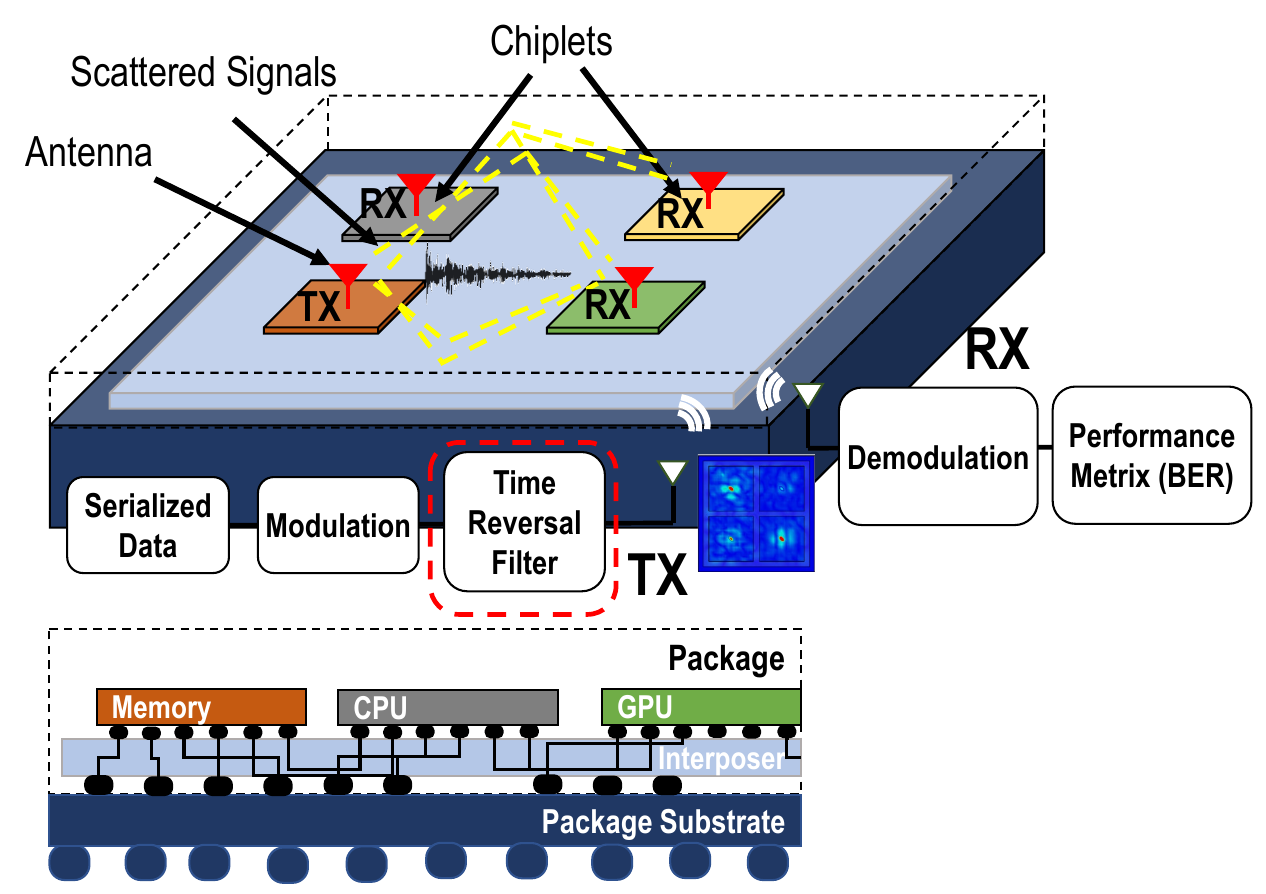}
\caption{Overview of the application of time reversal in a wireless network within package. Modulated data is transmitted through the proposed TR filter and focused on the intended receivers.}
\vspace{-0.4cm}
\label{fig:abstract}
\end{figure}

In previous work, we explored TR in a small set of chip configurations and frequencies \cite{bandara2023exploration,rodriguez2023}. In \cite{bandara2023exploration}, we proposed TR in a flip-chip structure at 60 GHz to mitigate the ISI and in \cite{rodriguez2023} we aimed for a TR approach for multiple channels at 140 GHz, in a multi-core architecture to mitigate CCI (Fig. \ref{fig:abstract}). However, in this past work, we considered an ideal TR filter. Perfect filters are impossible to fabricate in practice, and in this scenario, a high-resolution filter might be unfeasible due to area/power constraints. 

In this paper, we expand our work in \cite{bandara2023exploration,rodriguez2023} with a deep exploration of the effectiveness of TR in a broader set of settings and assess filters with a finite sampling rate. In particular, the contributions of this paper are as follows:

\begin{itemize}
    \item We evaluate the performance of wireless links in a four-chiplet computing system at 140 GHz (Fig. \ref{fig:abstract}), in single-link and multi-link settings. We observe order of magnitude improvements on the speed with TR, and reduced mutual coupling effects with multiple concurrent links.
    \item We assess the impact of changing the antenna operating frequency without modifying the environment. We observe that TR is compatible with a range of frequencies, without large variations in performance.
    \item We evaluate the impact of changing the environment, by modifying the number of chiplets and their size. We observe that these changes affect the performance of the wireless links.
    \item We consider filters with a finite sampling rate and measure the degradation of performance as the sampling rate is reduced. We observe gradual reductions in the achievable speeds due to  mismatch of sampling frequency on the TR filter with the ideal sampling rate.
\end{itemize}

The rest of the paper is organized as follows. In Section \ref{sec:background}, we provide some background related to the general problems of the on-chip scenario. In Section \ref{sec:system}, we formulate the system model and in Section \ref{sec:methodology}, we provide a detailed description of the methodology, including the simulation pipeline and the evaluated variations. We present the results obtained for the parametric characterization from the base scenario in Section \ref{sec:results1}, while Section \ref{sec:results2} evaluates the performance of TR with the changes in the antenna operating frequency and the geometry of the chip. Finally, in Section \ref{sec:results3}, we assess the trade-off between performance and feasibility of TR filters and  Section \ref{sec:Discussion} discuss the limitations. Section \ref{sec:conclusion} summarize the contribution and outlines directions for future work.

\section{Background}
\label{sec:background}

\subsection{Wireless Communications at the Chip Scale}

Wireless on-chip communications have been built with the foundation of using mmwave technology, as an overlay of wireless links on top of a wired NoC architecture. A variety of antennas and transceivers \cite{gutierrez2009chip, yeh2012design} have been proposed to be co-integrated with the processors, by considering the wireless channel inside the chip package as the propagation medium. WNoC complements NoC nicely with reconfigurability, low latency, and broadcasting capabilities for a high number of cores. In a dense wired routing network, WNoC can avoid unnecessary hops when communicating with distant cores.

To meet the expected performance of WNoC with high data rates and ultra-low latency, the wireless channels have to be shared among the antennas or the modulation order of data transmissions have to be increased. Therefore MAC protocols play an important role in maintaining reliable data communication, where several attempts have been made \cite{duraisamy2015enhancing, bandara2025, abadal2018medium} to adapt token-passing, CSMA to the chip environment by considering its unique requirements and characteristics. On the other hand, although the modulations are also crucial on enhancing the data rate, the area, power overhead of the chip scale transceivers and stringent BER requirements have limited the modulations to On-Off keying (OOK) \cite{yazdanpanah2022systematic, timoneda2020engineer}.

\subsection{Wireless Channel within the Chip Package}\label{sec2B}
The controlled materials and dimensions within the chip package let us tailor the wireless channel and know the nature of the medium beforehand \cite{Matolak2013CHANNEL,chen2019channel}. A computer chip is created by stacking layers of lossy silicon, metal, and dielectric \cite{markish2015chip}. In particular, the structure of the flip-chip configurations widely used in nowadays processors is, from top to bottom: a heat sink on top, the silicon substrate, a thin dielectric layer for the interconnections, and a layer of solder bumps that connect the chip to the PCB \cite{wright2006characterization}. 
Recently, a trend in computer architecture consists in the interconnection of multiple flipped chips in a co-planar configuration. To this end, the chips are connected through a silicon interposer with a much higher connection pin density than via a PCB \cite{zhang2015heterogeneous, kannan2015enabling}. All the work developed in this paper considers such a 2.5D integration scheme.

Regardless of the number of interconnected chips, communicating within the chip package with EM radiation has its unique characteristics. A large amount of the transmitted energy will be reflected by the metal enclosure, the edges, and obstacles in the structure, leading to a large delay spread with multipath propagation while creating a distinctive CIR between each transmitter and receiver \cite{Matolak2013CHANNEL}. Fortunately, the structure is designed beforehand, and all materials, nodes, and terminals will become fixed and known. This offers a significant advantage since the wireless channel can be treated as quasi-deterministic and time-invariant which enables a pre-characterization of the CIR that will not change and that requires a single estimation process and can be re-used moving forward.

\subsection{Time Reversal for On-Chip Communications}

\begin{figure}[!t]
\centering
\vspace{-0.2cm}
\includegraphics[width=\columnwidth]{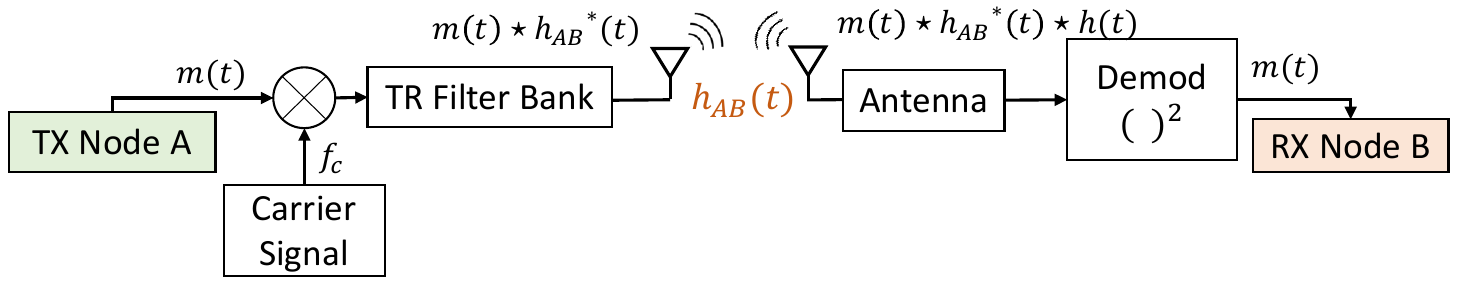}
\caption{Block diagram representing transmitting and reception with TR filter.}
\label{fig:trblock}
\vspace{-0.2cm}
\end{figure}

TR is a technique that can be used to compensate for the multipath effects that occur in complex transmission mediums with the use of adaptive energy focusing on the pre-characterized CIRs. The phase conjugate of the CIR is obtained, pre-emphasized with the transmitting symbols, and propagated through the same wireless medium such that the radiated energy is focused in both time and space at a target receiver as shown in Fig. \ref{fig:trblock}. TR acts as an ideal spatial matched filter with the process of auto-correlation of the transmitted time-reversed channel response with an equal wireless medium. 
As the chip-scale wireless communications are conducted in quasi-deterministic channels, TR is a promising technology to adapt as a solution to mitigate the ISI effects that occur due to energy reverberation. With prior known wireless channels, repetitive CIR characterization is not required.

Channel correlation plays a crucial role in TR processing. Even though TR was initially adapted for on-chip communications to mitigate ISI, later it was proved that TR could be used to reduce the CCI in same frequency/time signal transmissions \cite{bandara2023exploration,rodriguez2023}. The dispersive nature of the wireless medium inside the chip create unique wireless channels for each link. When these communicating channels are less correlated TR is capable of reducing the received energy to the nearby nodes, with the destructive interference occur due to low cross-correlation in between the channels. In return, it enables the wireless medium to have parallel concurrent signal transmissions by sharing the same bandwidth, thus improving the aggregate data rate of the system. However, this phenomenon is adverse with symmetric channels where the CCI could be high. In such cases, the known CIR can be analyzed before the transmissions to optimize the communication scheme. Nevertheless, the symmetric channels can be highly useful to focus the energy on multiple nodes with single transmitter excitation to enhance the broadcasting capabilities.

\section{System Model} 
\label{sec:system}
In this section, we provide a detailed description of the TR process and its influence on the transmission. The model is detailed for different scenarios involving one or more concurrent TR links.

\subsection{TR with a Single Receiver}\label{sec3a}

We analyze the case of TR-based energy focusing with respect to a single link. Let us consider, for the purpose of characterisation, that an impulse is excited from an arbitrary transmitter $i$ to an arbitrary receiver $k$. The characterized CIR is recorded and used to transmit modulated information via the channel $h_{ik}$, which is given by
\begin{equation}\label{eq1}
    h_{ik}(t)= \sum_{n=1}^{N} A_{n}e^{j\theta_n}\delta(t-\tau_n)\, ,
\end{equation}
where $\{A_{n},\theta_{n},\tau_{n}\}$ are amplitude, phase, and propagation delay of the multipath component $n$ of the CIR $h_{ik}$.

For simplicity, impulse radio OOK modulated symbols are assumed. Hence, symbols $m_{ik} \in \{1,0\}$, are sent through the characterized time-reversed filter and transported towards receiver $k$ from the transmitter $i$. The transmitted signal can be expressed as
\begin{equation}\label{eq2}
    s_{ik}(t)=h_{ik}^{*}(-t) \star m_{ik}  
\end{equation}
where $(\cdot)^{*}$ represents the complex conjugate operator and $\star$ indicates the convolution operator respectively.

The resultant energy concentration resembles the autocorrelation of the CIR as
\begin{equation}\label{eq3}
\begin{split}
    y_{ik}(t)&=\{s_{ik}(t) \} \star h_{ik}+n_{k}\\
             &=m_{ik} \star h_{ik}^{*}(-t) \star h_{ik}+n_{k}\\
             &=m_{ik} \star R_{ik}^{A}(t) +n_{k}\\
\end{split}    
\end{equation}
where $R_{ik}^{A}(t)$ denotes the autocorrelation function of the channel $h_{ik}$ and $n_{k}$ is the Additive White Gaussian Noise (AWGN) component at the receiver $k$. With an ideal TR filter response, the autocorrelation property eliminates the multipath effects with constructive and destructive interference cancellation and gains strong temporal focusing while improving the signal-to-interference plus noise ratio (SINR) at the respective receiver. Yet, the cross-correlation between the transmitter-receiver channel and the channels at nearby locations will affect the degree of spatial focusing and residual side lobes may persist.

Next, we assume a case where the transmitted signal $s_{ik}$ is propagated towards an off-target receiver $k'$. Then,
\begin{equation}\label{eq4}
\begin{split}
    y_{ik'}(t)&=\{s_{ik}(t) \} \star h_{ik'}+n_{k'}\\
             &=m_{ik} \star R_{ikk'}^{C}(t) +n_{k'}\\
\end{split}    
\end{equation}
where $R_{ikk'}^{C}(t)$ represents the cross correlation function in between the arbitrary channels $h_{ik}$ and $h_{ik'}$. 
For further analysis of the impact of channel correlation functions on TR, let us consider the convolution of the CIR $h_{ik'}$ with the time-reversed $h_{ik}^{*}$:
\begin{equation}\label{eq5}
\begin{split}
 h_{ik'}(t)\star h_{ik}^{*}(-t)&=\int_{-\infty}^{\infty} h_{ik'}(\tau) h_{ik}^{*}(-(t-\tau)) d\tau\\
\end{split}    
\end{equation}

By substituting Equation (\ref{eq1}) in Equation (\ref{eq5}), the convolution operation can be obtained as
\begin{equation}\label{eq6}
\begin{split}
 h_{ik'}(t)\star h_{ik}^{*}(-t)&=\int_{-\infty}^{\infty} \sum_{n'=1}^{N' }A_{n'}e^{j\theta_{n'}}\delta(\tau-\tau_{n'})\\&\sum_{n=1}^{N} A_{n}e^{-j\theta_n}\delta(-(t-\tau)-\tau_n) d\tau\\
\end{split}    
\end{equation}
where $\{A_{n'},\theta_{n'},\tau_{n'}\}$ are amplitude, phase, and propagation delay of the multipath component $n'$ of the CIR $h_{ik'}$.

Thus the simplified cross-correlation expression can be obtained as
\begin{equation}\label{eq7}
\begin{split}
 R_{ikk'}^{C}(t)&= \sum_{n'=1}^{N' }\sum_{n=1}^{N} A_{n'}A_{n}e^{j(\theta_{n'}-\theta_{n})}\delta(t-(\tau_{n'}+\tau_{n}))
\end{split}    
\end{equation}

With the above expression, it is observed that when the transport signal is passed through symmetric channels (auto correlation index is high) the energy components are constructively added with a null phase shift while having the maximum energy concentration at the point of maximum propagation delay; $max(\tau_{n})$.  When the TR adopted for dissimilar CIR, the cross-correlation index is phase shifted and the energy concentration will be controlled by the shifted amount of phase.

\subsection{TR with Multiple Receivers}\label{sec3b}

Subsequently, let us consider TR precoded transmission with multiple receivers. In the cases that follow, suppose there are $N_t$ transmitters with $N_r$ receivers simultaneously communicating with each other on an interposer package.

This scenario resembles a parallel information shared in between cores, to improve the aggregate data rate of the system. Here, each $i^{th}$ transmitter wishes to transmit the TR precoded modulated information to the intended receiver $k$. Thus, the transmitted signal can be obtained as
\begin{equation}\label{eq10}
s^{r}_{ik}(t)=  \sum^{N_t}_{i^{'}=1,i^{'} \neq i}h_{i^{'}k}^{*}(-t)\star m_{i^{'}k}+ h_{ik}^{*}(-t)\star   m_{ik}
\end{equation}

TR precoding is done based on the simultaneous transmissions utilized in the particular transmitter to the intended receivers. Note that according to the receiver selection, other TR filters based on the pre-characterized CIRs of the wireless links can be assumed as zero.

The received signal at a single $k^{th}$ receiver can be expressed as
\begin{equation}\label{eq11}
\begin{split}
    y^{r}_{ik}(t)&=s^{r}_{ik}(t) \star h_{ik}+n_{k}\\
             &=\bigg\{\sum^{N_t}_{i^{'}=1,i^{'} \neq i}h_{i^{'}k}^{*}(-t)\star m_{i^{'}k}+ h_{ik}^{*}(-t)\star   m_{ik}\bigg \}\star h_{ik} \\& +n_{k}\\
            &=\underbrace{\sum^{N_t}_{i^{'}=1,i^{'} \neq i} R_{i'kk}^{C}(t)\star m_{i'k}}_\text{Interference}+ \underbrace{  R_{ik}^{A}(t)\star m_{ik} }_\text{Signal}+n_{k}\\
\end{split}    
\end{equation}

As we observe in Equation (\ref{eq11}), the interference is dependent on the channel cross-relation between the focused channel CIR with the interference channels (refer to Fig. \ref{fig:cases}) due to the non-orthogonality of the signal transmissions.

\section{Methodology}
\label{sec:methodology}
\begin{figure}[!t]
\centering
\vspace{-0.2cm}
\includegraphics[width=0.8\columnwidth]{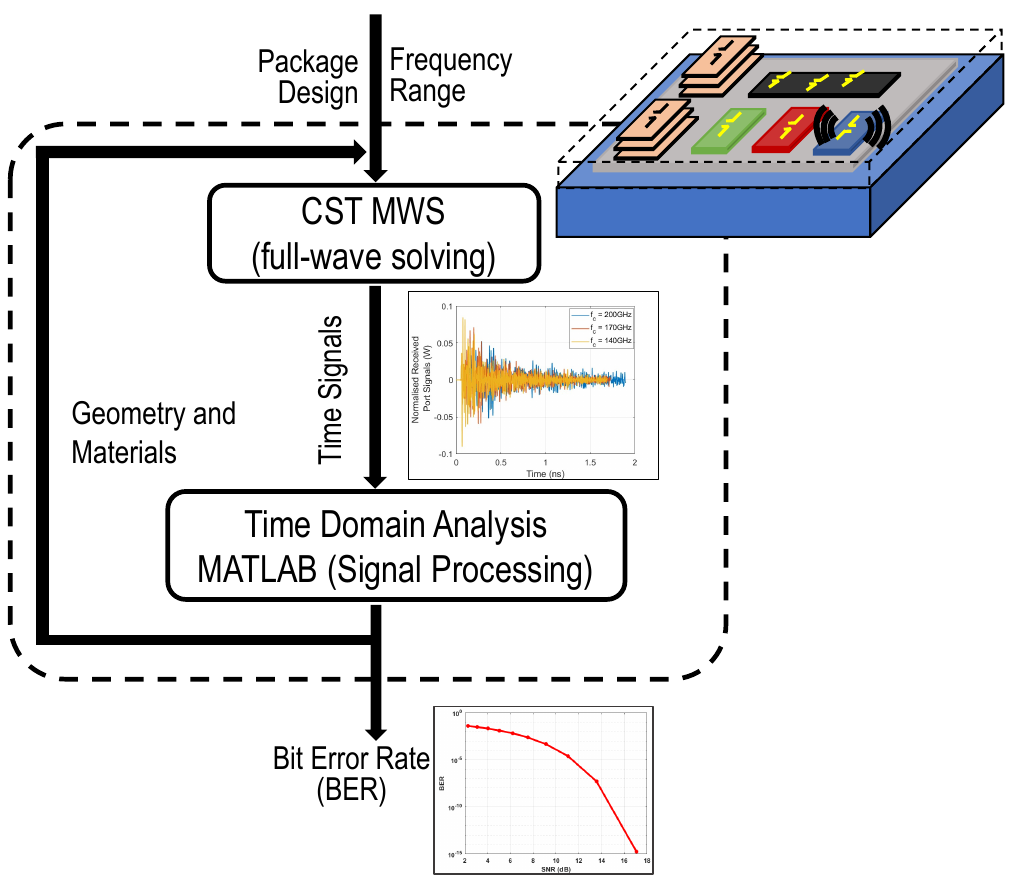}
\caption{Summary of the proposed methodology: Characterization of channel impulse response and analyzing the proposed communication with MATLAB.}
\vspace{-0.2cm}
\label{fig:msummary}
\end{figure}

The overview of the proposed methodology is shown in Fig. \ref{fig:msummary}. We follow a sequential methodology for this work to (i) systematically pre-characterize the wireless channel with CIR by using CST Microwave Studio \cite{CST} and (ii) model the communication system on MATLAB to analyze the performance of the TR technique on modulated streams of data. We detail them next.

\subsection{Full-wave Simulation}
The simulations made in this work are conducted in an interposer. The detailed configuration of the flip-chip layers is seen in Table \ref{tab:flipParams}. To create the interposer we placed a number of flip-chip structures on top of a stack of silicon interposer and two layers of copper. The space between chiplets is filled with vacuum and the configuration is completely enclosed by a Perfect Electric Conductor (PEC). Each chiplet is augmented with a number of antennas placed in symmetrical positions in a Trough-Silicon-Vias (TSV) approach \cite{bandara2023exploration} as vertical monopoles. Unless noted, we consider four chiplets with three antennas per chiplet tuned at a frequency of 140 GHz.

\begin{table}[!t] 
\caption{Evaluated Structure and Dimensions.} 
\label{tab:flipParams}  
\vspace{-0.1cm}
\centering
\begin{tabular}{ccccccc} 
\hline
{\bf Parameter} & {\bf Thickness} & {\bf Materials} & {\bf Units} \\
\hline
Lateral Spacer & - & Vacuum & N/A\\
Heat Spreader & 0.5 & Aluminum Nitride & mm \\
Silicon die & 0.5 & High-res. Silicon & mm \\
Chiplet insulator & 0.01 & SiO\textsubscript{2} & mm \\
Bumps & 0.0875  & Copper &  mm\\
\hline
\end{tabular}
\end{table}

\begin{table}[!t] 
\caption{Evaluation Scenarios.} 
\label{tab:flipVar}  
\vspace{-0.1cm}
\centering
\begin{tabular}{c|cc|cc} 
\hline
{\bf Variables} & {\bf Default} & {\bf Section} & {\bf Variations} & {\bf Section} \\
\hline
Number of chiplets & 4  & Sec. V & 4--16 & Sec. VI-A \\
Frequency (GHz) & 140 & Sec. V  & 140--200 & Sec. VI-B \\
Sampling Rate (GHz) & Ideal & Sec. V & 80--400 & Sec. VII \\ 
\hline
\end{tabular}
\end{table}

Moving forward, we performed several simulations to study the TR technique, its behavior, and application possibilities for changes in the base scenario in terms of frequency of operation and changes in the structure's geometry. We refer the reader to Table \ref{tab:flipVar} for details on the evaluated variations.

\begin{figure}[!t]
\centering
\vspace{-0.2cm}
\includegraphics[width=0.9\columnwidth]{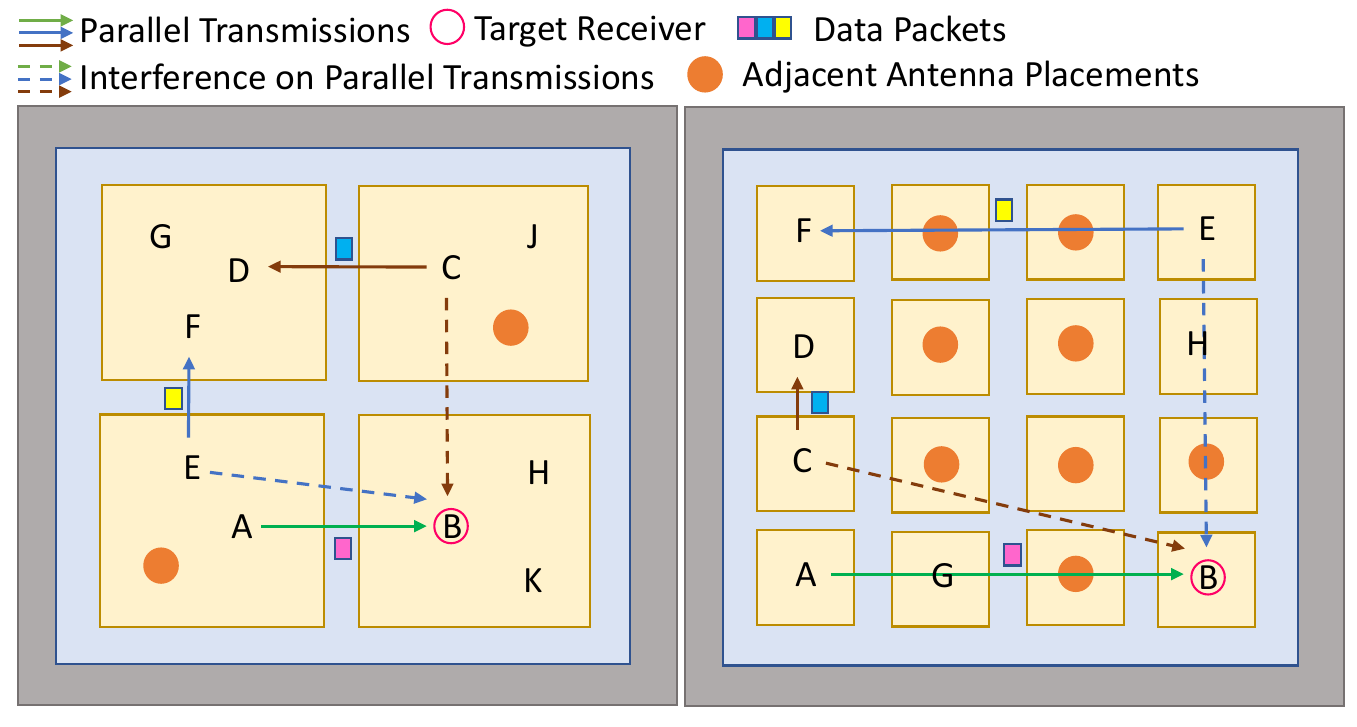}
\caption{Evaluated TR scenarios for four and sixteen chiplets in an interposer package, with multiple concurrent transmitters at the same time, different data to different receivers.}
\label{fig:cases}
\vspace{-0.2cm}
\end{figure}

The structures where the parametric sweep takes place are seen in Fig. \ref{fig:cases}. We illustrate the location of the antennas and the links used for measurement of the quality of transmission. As observed, we change the number of chiplets to study the impact of the layout on the effectiveness of TR. In particular, we introduce more chips in the same amount of space, hence decreasing the chiplet size and only placing one antenna per chiplet.

\begin{figure}[!t]
\centering
\includegraphics[width=0.3\textwidth]{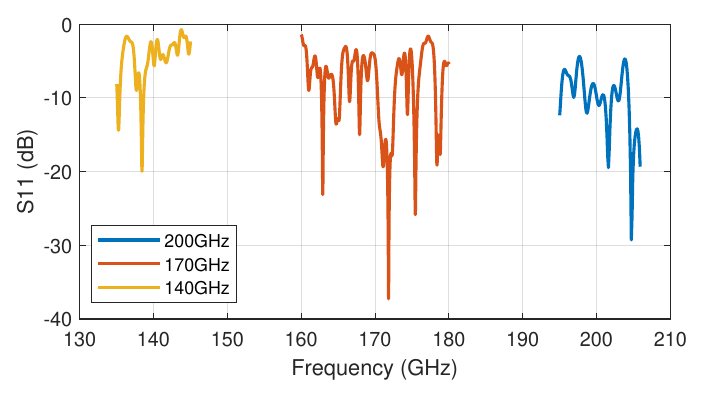}
\caption{Return loss for antennas at 140/170/200 GHz, four chiplets.}\label{fig:s11}
\end{figure}

Each scenario is simulated in CST Microwave Studio using the time-domain solver. To study the dependence of TR on the frequency used for communication, we adjust the length of the vertical monopole antennas to resonate around 140, 170, and 200 GHz (Fig. \ref{fig:s11}). We can also see several notches, which suggest that we are indeed in a reverberant scenario; the antenna can couple with nearby and distant elements, changing the impedance and hence mismatching the antenna. As a result, we also adapt the length of the antenna when varying the dimensions of the chiplet.

Once the antennas are modeled correctly, we apply an impulse to the transmitting antenna to obtain the CIR at the output. Hence, the outcome of the simulations is the CIR from a given port to all others and we obtain channel data to evaluate any wireless link we consider for transmission. The distance in between the antennas is in the range between 1.29$\lambda$-4.38$\lambda$. Since all antennas are included in the simulations during the channel pre-characterization phase, we guarantee that the mutual coupling effects of the short-distanced antenna placements are captured by the CIR. 

Finally, we note that, to validate the TR approach, we record the CIR and excite the transmitting antenna with the reversed version of the CIR. This simulation is used to obtain the field distribution and see the spatiotemporal focusing, as well as to later compare the results with that of a simple convolution between the filter and the actual channel in MATLAB.

\subsection{PHY Layer Simulation}
As the next step, the characterized CIR is used to process the proposed communication scheme with TR. 
First, we generate a pseudo-random string of bits with the preferred data rate. The data is then modulated with OOK modulation to create a train of impulses, although the methodology is easily applicable to continuous-wave modulations, in which case the data would be multiplied by a carrier wave with a frequency matching the antenna resonance frequency. 

In transmission, the modulated symbols are convoluted with the TR filters and transported in the wireless channel. AWGN noise is added to the transport signal and the noise power is modeled as $P_{AWGN}=KTB$, where $K$ is the Boltzmann constant, $T$ is the chip temperature and $B$ is bandwidth. The cross-interference with respect to each signal transmission is obtained as in Equation (\ref{eq11}). 

The TR filter is initially modeled as the ideal (continuous) time-reversed CIR perfectly matching the characterized channel. However, in Section \ref{sec:results3}, we also perform a gradual degradation of the filter sampling rate in the default simulated scenario to consider a more realistic filtering approach and observe the resulting degradation in performance of the wireless link.

At the receiver, energy detection is implemented by integrating the received signals over the respective time intervals based on the bandwidth. The computed energy values are compared with a threshold to decode the data stream. The threshold value is decided a posteriori and on a per-link basis to minimize the BER. The BER values represent pre-Forward Error Correction (FEC) performance of the links.
 
\subsection{SINR Analysis with TR}\label{sec3c}

To quantify the impact of TR on the effectiveness of the signal transmission, the SINR shall be assessed for both single-link and multi-link TR precoded communications.

We detect the peak energy concentration inside a defined time window with general energy detection \cite{lehtomaki2005analysis}. The length of the time window ($t_{rate}$) is obtained based on the expected bandwidth of the signal transmission. The energy is measured by placing the time window such that the maximum energy concentration occurs at $t_{rate}/2$ and the energy components outside the defined time duration is considered as interference due to multipath effects.

The SINR is defined for two cases as aforementioned in Section \ref{sec3a} and Section \ref{sec3b}. Therefore, SINR for single transmitter single receiver channel can be obtained as
\begin{equation}\label{eq12}
\begin{split}
SINR_{stsr}=\frac{\int^{t_2}_{t_1} y_{stsr}(t)^2 dt }{I_{stsr} + n_{k}}
\end{split}    
\end{equation}
where $ y_{stsr}(t)=m_{ik}\star R^{A}_{ik}(t)$, $I_{stsr}=\int^{t_1}_{0} y_{stsr}(t)^2 dt+\int^{2max(\tau_{n})}_{t_2}y_{stsr}(t)^2dt $, $t_1=max(\tau_n)-t_{rate}/2$, $t_2=max(\tau_n)+t_{rate}/2$. Here, $y_{stsr}(t)$ represents the signal energy of the received signal, while $I_{stsr}(t)$ denotes the ISI caused due to multi path effects of the channel.

Similarly, with the received signal in Equation (\ref{eq11}), the SINR for multiple transmitters and multiple receivers can be expressed as below.

\begin{equation}\label{eq14}
\begin{split}
SINR_{mtmr}=\frac{\int^{t_2}_{t_1} y_{mtmr}(t)^2 dt }{I_{mtmr} + I^{r}_{CCI}+n_{k}}
\end{split}    
\end{equation}
where $ y_{mtmr}(t)=m_{ik}\star R^{A}_{ik}(t)$, $I_{mtmr}=\int^{t_1}_{0} y_{mtmr}(t)^2 dt+\int^{max(\tau_{n}+\tau_{n'})}_{t_2}y_{mtmr}(t)^2dt$, $ I^{r}_{CCI}=\int^{t_2}_{t_1} (R_{i'kk}^{C}(t)\star m_{i'k})^2 dt$. $y_{mtmr}(t)$ denotes the energy of the received signal, $I_{mtmr}(t)$ captures the ISI due to multi path propagation and $I_{CCI}(t)$ is the CCI due to multiple simultaneous transmissions.

\section{Time Reversal in a Multi-Chip Scenario}
\label{sec:results1}
For the baseline scenario, the simulations have been executed for several transmission cases in a four-chiplet system at 140 GHz and assuming ideal filters. Also, the links contemplated are within the same chip and among different chips. The results entail single-link and multiple-link transmissions. Within the multiple links scenario, we made simulations for multiple transmitters and multiple receivers.

\subsection{Impact of Single-link Transmissions}

\begin{figure*}[!t]
\centering
\begin{subfigure}[t]{0.33\textwidth}
\includegraphics[width=0.47\textwidth]{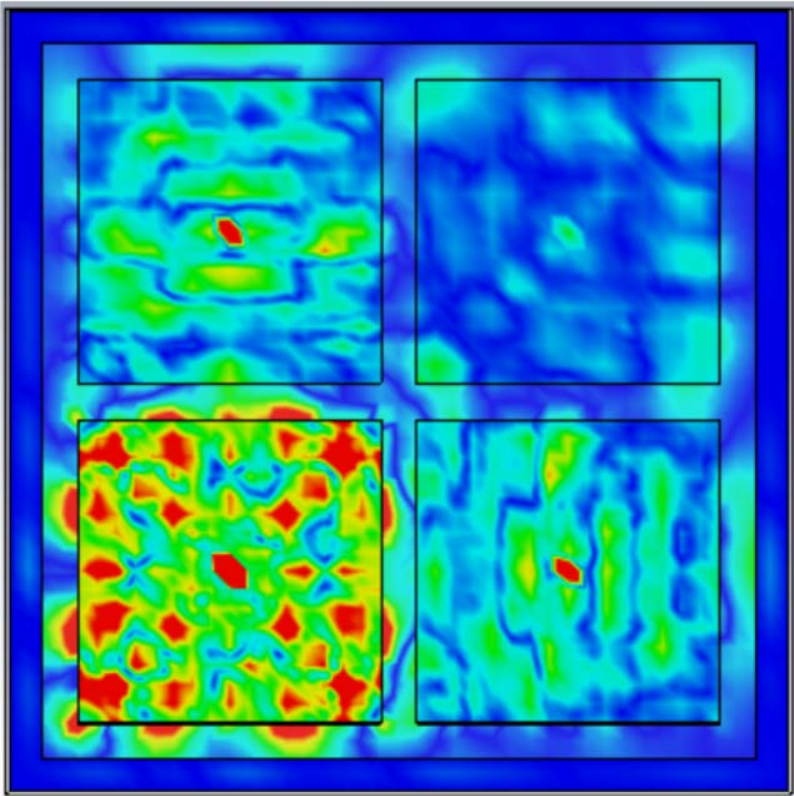} 
\includegraphics[width=0.47\textwidth]{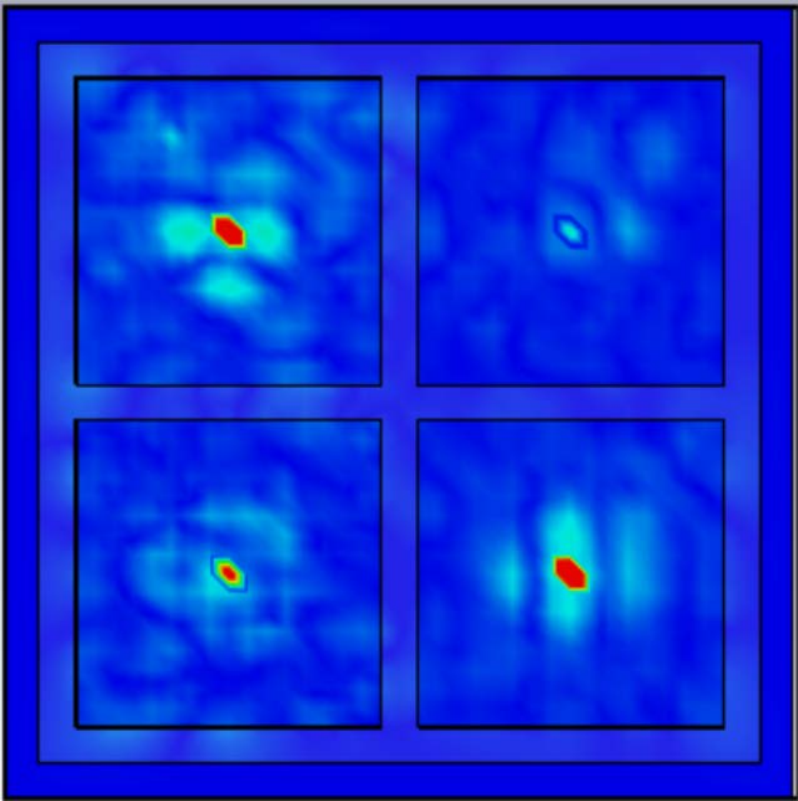}
\caption{}
\label{fig:singlelinka}
\end{subfigure}
\begin{subfigure}[t]{0.22\textwidth}
\includegraphics[width=\textwidth]{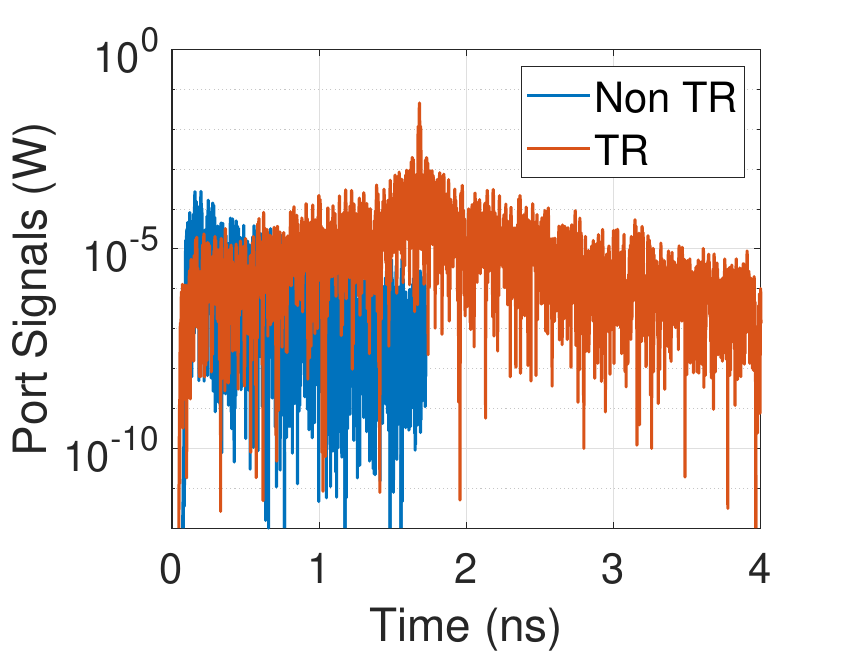}
\caption{}
\label{fig:singlelinkb}
\end{subfigure}
\begin{subfigure}[t]{0.2\textwidth}
\includegraphics[width=\textwidth]{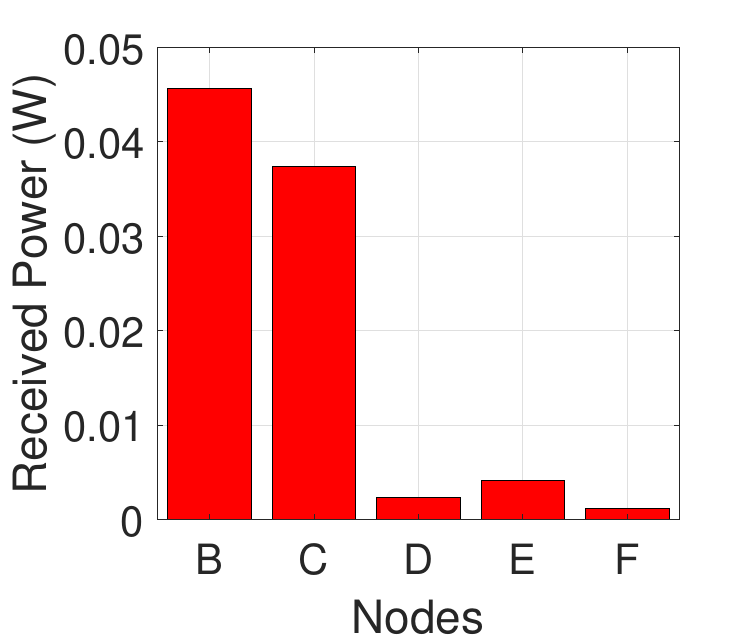}
\caption{}
\label{fig:singlelinkc}
\end{subfigure}
\begin{subfigure}[t]{0.23\textwidth}
\includegraphics[width=\textwidth]{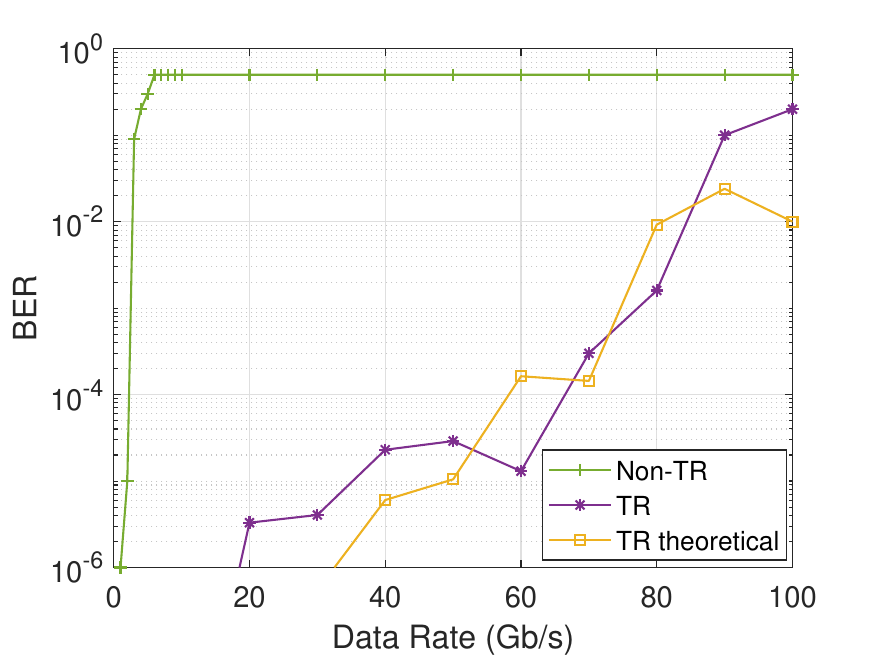}
\caption{}
\label{fig:singlelinkd}
\end{subfigure}
\vspace{-0.2cm}
\caption{TR with single-link transmissions in an interposer package at 140GHz (a) Spatial distribution of field before and applying TR. (b) Signal received over time in node $B$ before and after applying TR. (c) Total received power in the intended receiver $B$ and non-intended receivers $C$--$F$ (d) BER as a function of data rate for a total power of 10 dBm in the TR and Non-TR cases. }
\label{fig:singlelink}
\vspace{-0.4cm}
\end{figure*}

Fig. \ref{fig:singlelink} shows a summary of the results of TR obtained with our end-to-end methodology. Fig. \ref{fig:singlelinka} illustrates the spatiotemporal focusing of the technique (right) in comparison with normal transmission (left). We can see the cleanness of the TR transmission with high concentration in targeted spots and low concentration elsewhere, except for symmetric points. On the other hand, the electrical field in the non-TR case is scattered, with some concentration in the desired target but surrounded by large amounts of interference, which tarnishes any concurrent transmission attempt. 

As we further show in Fig. \ref{fig:singlelinkb}, the TR signal at the receiver is focused in the time domain, showing a large and very concentrated peak of 45 mW; whereas the non-TR signal appears spread and with the peak only reaching 0.2 mW. 

With respect to CCI, an important conclusion can be drawn from Fig. \ref{fig:singlelinkc}. This image shows the ability of TR to compress the energy concentration in non-desired targets if and when the CIR of each wireless path is not correlated with the targeted receiver. In this particular case, node $C$ is placed in a relatively symmetric position with respect to the targeted node $B$ and, hence, receives a relatively large interference. An opposite behavior is observed for nodes $D$ to $F$, for which the interference is largely suppressed.

In Fig. \ref{fig:singlelinkd} we compare the BER on TR and non-TR single-link transmissions. Theoretical BER with Q-function for OOK modulation was compared with simulations. It is observed that with the superior spatial-temporal focusing in high data rates TR shows a better link quality by reducing the ISI. However in non-TR transmissions, the BER is increased when the data rate is reaching to a few Gb/s with the high delay spread of the channel as observed in Fig. \ref{fig:resonancefrequencya}.

\subsection{Impact of Multi-link Transmissions}

\begin{figure*}[t]
\centering
\begin{subfigure}[t]{0.3\textwidth}
\includegraphics[width=\textwidth]{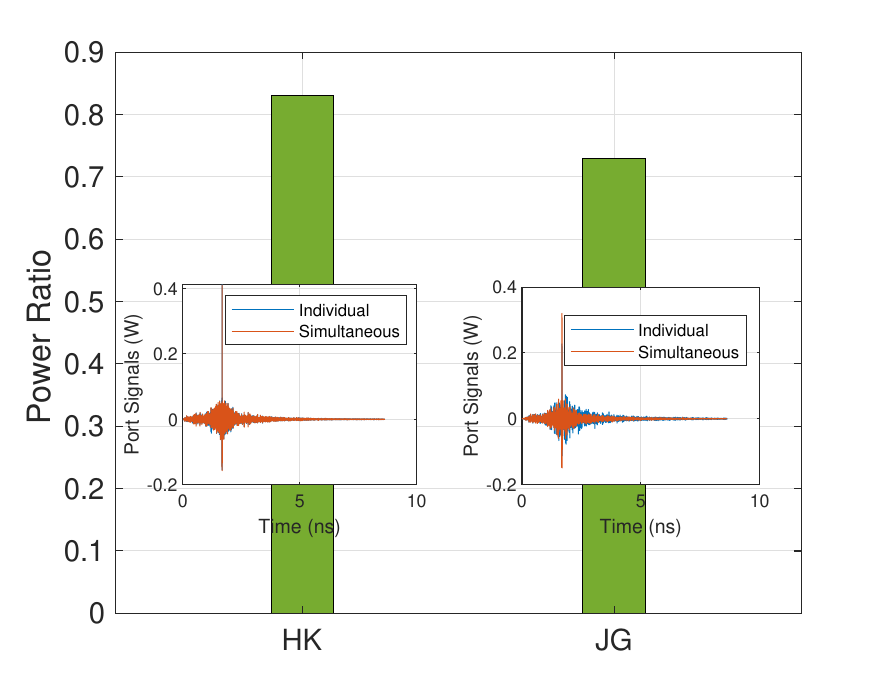}
\caption{}
\label{fig:multilinka}
\end{subfigure}
\hspace{-0.4cm}
\begin{subfigure}[t]{0.3\textwidth}
\includegraphics[width=\textwidth]{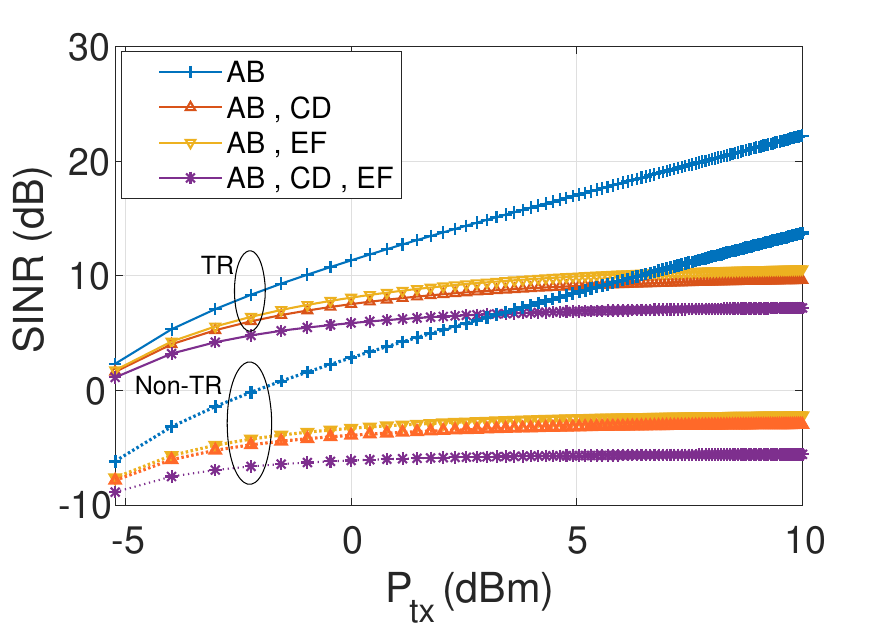}
\caption{}
\label{fig:multilinkb}
\end{subfigure}
\begin{subfigure}[t]{0.3\textwidth}
\includegraphics[width=\textwidth]{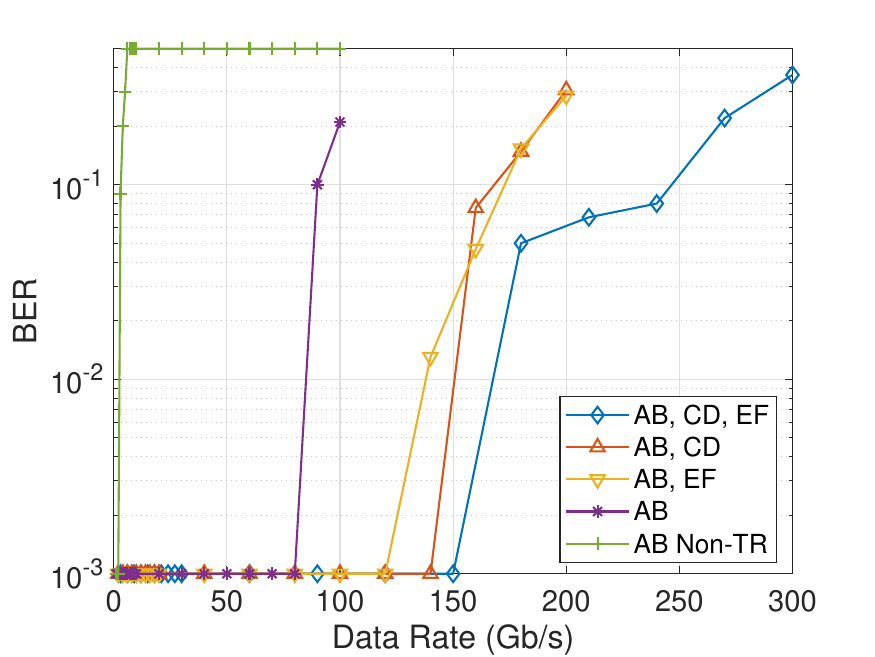}
\caption{}
\label{fig:multilinkc}
\end{subfigure}
\vspace{-0.2cm}
\caption{Impact of multi-link transmissions in four-chiplet case at 140GHz (a)  Measurement of mutual coupling effects with individual and simultaneous TR transmissions for $HK$ and $JG$ links  (b) SINR for non-TR and TR links as a function of transmitted power  (c) BER as a function of the aggregated data rate for a total power of 10 dBm in the TR and Non-TR cases. }
\vspace{-0.2cm}
\label{fig:multilink}
\end{figure*}

To evaluate TR with parallel transmissions, first we assess the near-field coupling effects of the closely placed monopole antennas as shown in Fig. \ref{fig:cases}. In terms of mutual coupling, even though the antennas are located in a close proximity to each other, we observe in Fig. \ref{fig:multilinka} the spatiotemporal focusing shown with TR in simultaneous transmission is similar with sequential individual TR transmissions. In here, we measure the ratio of average received energy at nodes $G$ and $H$, with respect to the single link TR transmissions and multi-link signal transmissions from the nodes $J$ and $K$.
Thus, we observe that with the energy ratios the spatial isolation offer with TR precoded signals enable us to  transmit multiple signals from multiple receivers at the same time, despite of the near-field impairments. 

The energy focusing with TR precoded multi-link transmissions were measured with respect to received signal strength with a total of 10 dBm transmit power in Fig. \ref{fig:multilinkb}. The SINR of a single-transmitter-single-receiver and multiple-transmitter-multiple-receiver cases, as mentioned in Equation (\ref{eq12}), Equation (\ref{eq14}) were obtained respectively. It was observed that, when the transmission power is swept, an 8 dB SINR gap in between TR and non-TR signal transmissions appears. Based on the positions and the distance between the links CD and EF, there is approximately 1 dB gap in the SINR. However, it is illustrated that, in all cases, transmissions with TR have high SINR with focused signal and reduced multipath effects. In summary, these results are indicative of the capacities of TR.

Furthermore, the performance of TR with multiple transmissions were measured in terms of BER as a function of the aggregated data rate (i.e. the sum of the data rates of all parallel transmissions), as illustrated in Fig. \ref{fig:multilinkc}. It was observed that even with three concurrent communications, the focused energy is higher, and thus the BER is improved at the receiver. In terms of data rate, a significantly high aggregate data rate can be obtained with time reversal. In particular, a speed of 50 Gb/s is obtained with an error rate below 0.001 with a single-link transmission. Compared with the non-TR transmission, the TR case achieves an order of magnitude higher speed for the same BER. Further, as we add new concurrent links to the simulation, we observe that the aggregate data rate increases with the number of deployed, although the system starts to saturate with three concurrent links in this particular case.


\section{Sensitivity Analysis}
\label{sec:results2}


Next, we want to determine the limits of our proposed technique and explore possible shortcomings that arise from changes in the structure as well as variations with the frequency.

\begin{figure*}[!t]
\centering
\begin{subfigure}[t]{0.24\textwidth}
\includegraphics[width=\textwidth]{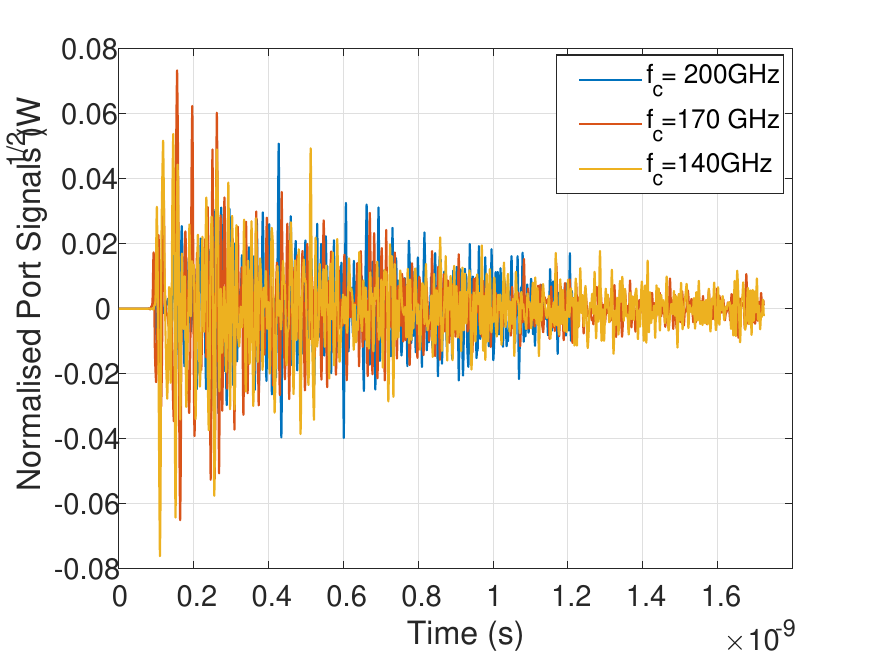}
\caption{}
\label{fig:resonancefrequencya}
\end{subfigure}
\hspace{-0.4cm}
\begin{subfigure}[t]{0.24\textwidth}
\includegraphics[width=\textwidth]{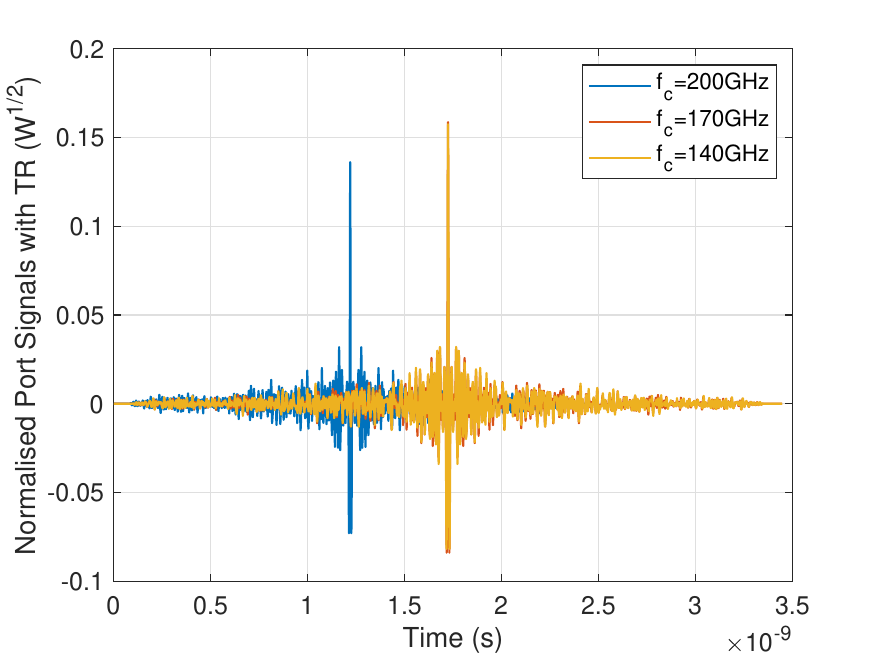}
\caption{}
\label{fig:resonancefrequencyb}
\end{subfigure}
\begin{subfigure}[t]{0.25\textwidth}
\includegraphics[width=\textwidth]{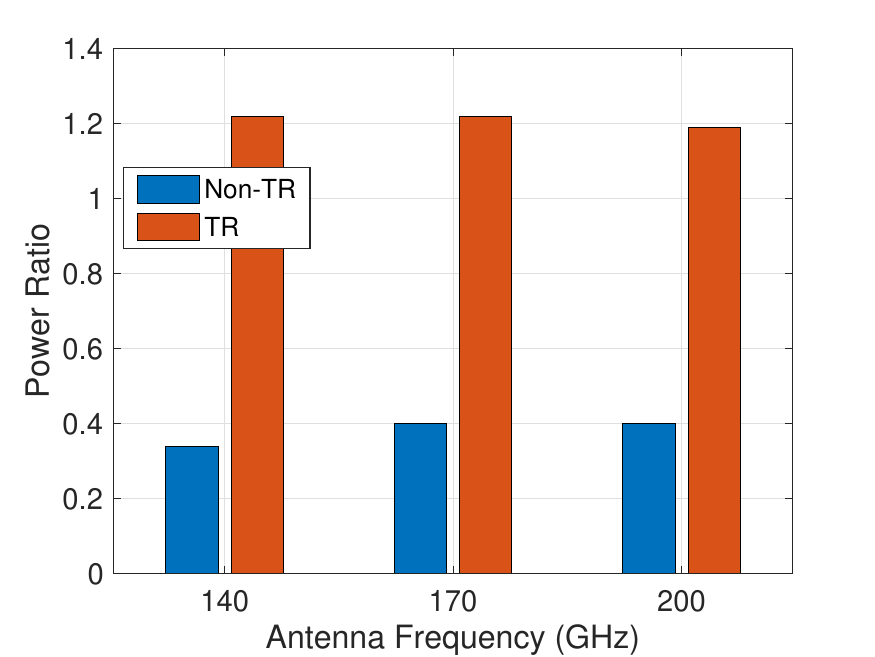}
\caption{}
\label{fig:resonancefrequencyc}
\end{subfigure}
\begin{subfigure}[t]{0.25\textwidth}
\includegraphics[width=\textwidth]{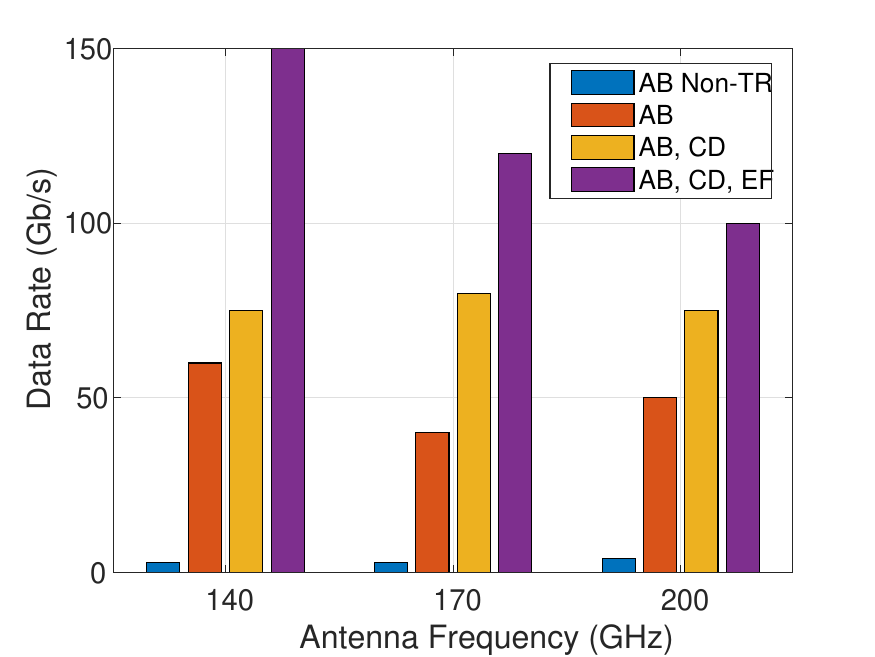}
\caption{}
\label{fig:resonancefrequencyd}
\end{subfigure}
\vspace{-0.2cm}
\caption{Impact of resonance frequency on validity of TR approach (a) CIR for the AB link without
TR and (b) with TR. (c) Power concentrated on the desired target node E divided by the power received at the rest of the terminals. (d) Aggregated data
rate with different techniques and number of concurrent links for BER of 0.1 and 0 dBm per transmitter.} 
\vspace{-0.4cm}
\label{fig:resonancefrequency}
\end{figure*}


\subsection{Impact of Communication Frequency}\label{sub1sec5}

As it is beneficial to count with multiple frequencies for transmission, the TR evaluation is made also for multiple frequencies, maintaining the rest of the interposer landscape intact. The peak energy concentration with TR on the focused receiver is based on two main factors; the delay spread and the received signal strength of the characterized CIR. When the center frequency of the antenna is increased, the delay spread generally tends to get shorter, and based on the scattered energy components on the chip package, the received signal strength on each antenna could be increased or decreased. This variation in frequency affects the receiver's peak energy concentration and thus the system's BER. We evaluate these effects next.


Fig. \ref{fig:resonancefrequencya} and Fig. \ref{fig:resonancefrequencyb} shows the general behavior of characterized CIR and the received peak energy concentration with different antenna resonance frequencies. It is observed that, even with a short delay spread, a clean focused energy peak is obtained at different studied frequencies. The specific peak instantaneous power and the width of the response depends on the length of the CIR and total energy of the CIR. While we observe some variations on these values, there is not an specific trend or large impact. This is because the scenario keeps being fairly reverberant at all frequencies, and the TR filter is adapted to each new measured CIR.



To assess the degree of spatial focusing of our technique, we evaluate the ratio of the power in the desired target to the power of the rest of ports of the system. As observed in Fig. \ref{fig:resonancefrequencyc} for a particular link, we observe how TR focuses the energy on the target destination, as the power is superior to the sum of the powers in all the other ports. In contrast, the non-TR case will probably lead to a large CCI, since the power at the intended destination is just a small fraction of the power arriving to all other ports. The difference between both cases is around 3$\times$ with small variation when changing the operation frequency of the wireless network. Hence, this confirms that TR could be used simultaneously to frequency multiplexing in a single package.

The TR capabilities have been further tested by obtaining the BER as a function of the data rate of an OOK modulated stream in all frequencies. 
The results, summarized in Fig. \ref{fig:resonancefrequencyd} for a fixed BER of 0.1, show a consistent behavior of the TR technique. In general, it is observed that TR works well at all frequencies, with a large gap between the non-TR and TR achievable rates. As the frequency increases, the performance of TR can increase or decrease depending on the particular shape of the channel at that particular link at the specific frequency, as well as the interference pattern it generates. Interestingly, the single-link performance decreases from 140 GHz to 170 GHz and then rises again. Where as the three-link performance decreases sharply when shifting from 140 GHz to higher frequencies.

\begin{figure*}[t]
\centering
\begin{subfigure}[t]{0.245\textwidth}
\includegraphics[width=\textwidth]{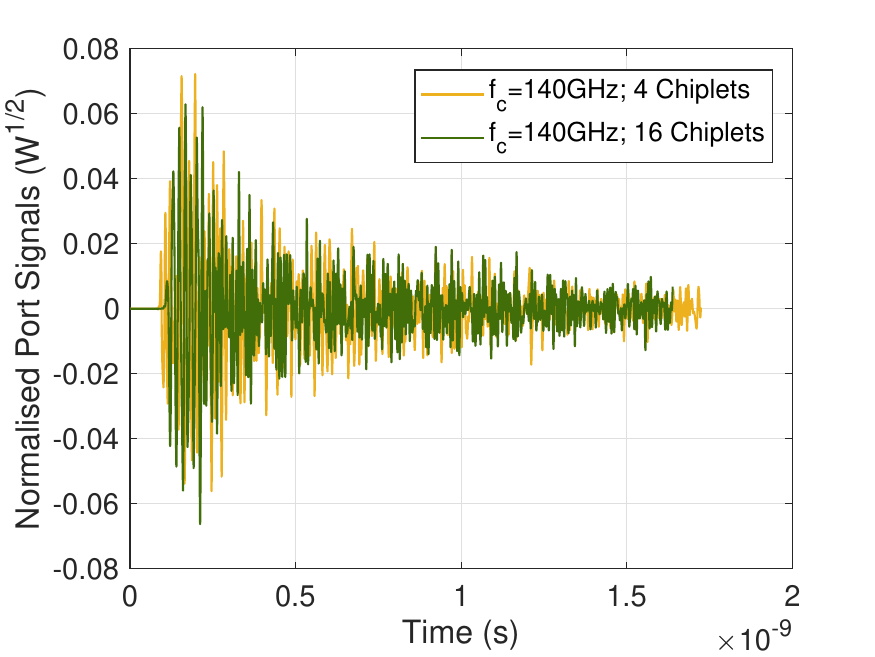}
\caption{}
\label{fig:numchipletsa}
\end{subfigure}
\hspace{-0.4cm}
\begin{subfigure}[t]{0.245\textwidth}
\includegraphics[width=\textwidth]{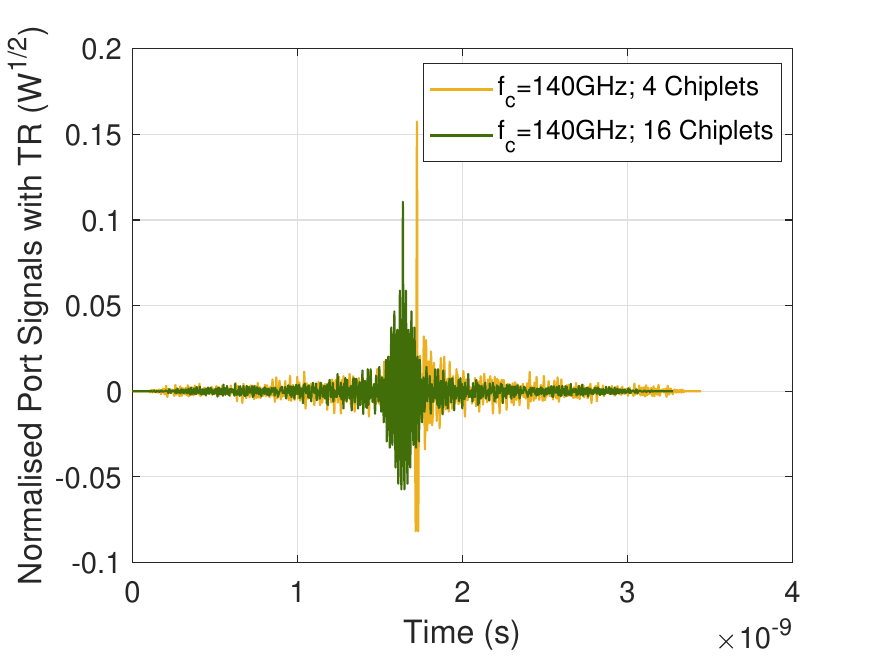}
\caption{}
\label{fig:numchipletsb}
\end{subfigure}
\begin{subfigure}[t]{0.245\textwidth}
\includegraphics[width=\textwidth]{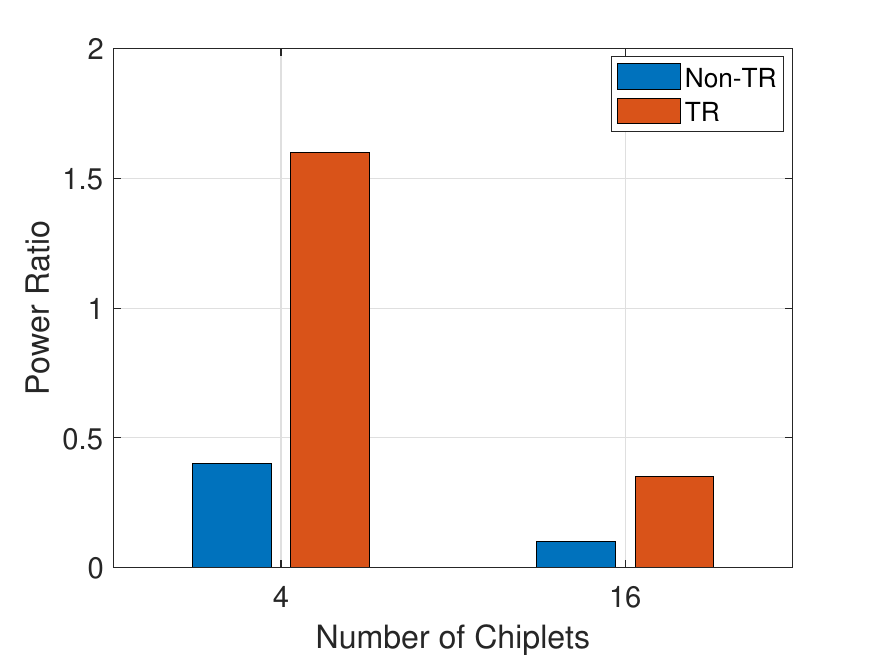}
\caption{}
\label{fig:numchipletsc}
\end{subfigure}
\begin{subfigure}[t]{0.245\textwidth}
\includegraphics[width=\textwidth]{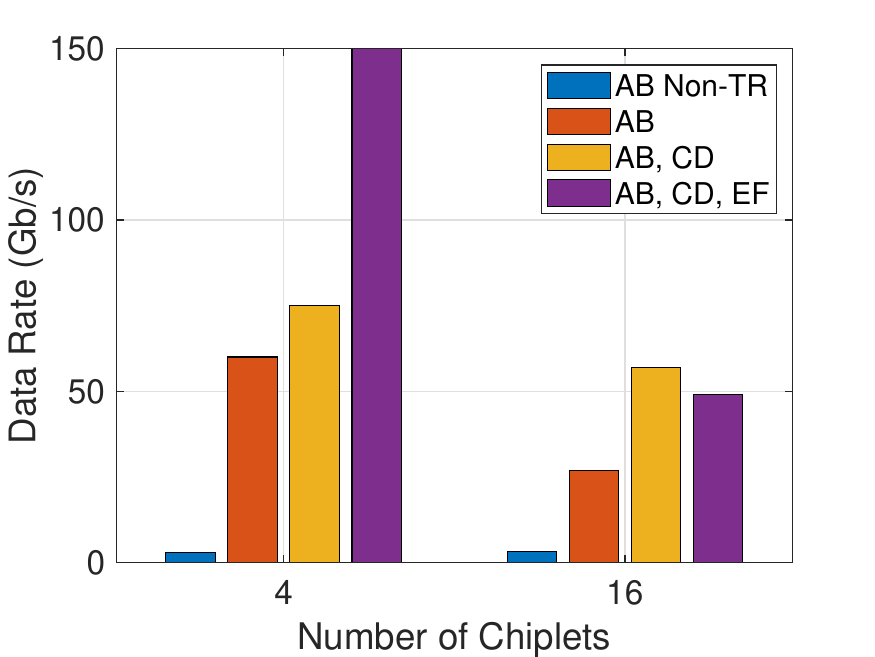}
\caption{}
\label{fig:numchipletsd}
\end{subfigure}
\vspace{-0.2cm}
\caption{Impact of the number of chiplets and their dimensions on the validity of the TR approach. (a)  CIR for the AB link without
TR and (b) with TR. (c) Power concentrated on the desired target to the power received by the rest of the terminals for links $GC$ and $HF$ in the four-chiplet and sixteen-chiplet case (d) Aggregate data rates for BER = 0.1 and 0 dBm per transmission.}
\vspace{-0.4cm}
\label{fig:numchiplets}
\end{figure*}

\subsection{Impact of Number of Chiplets}
Next, we evaluate the TR technique in a 16 chiplets interposer, as illustrated in Fig. \ref{fig:cases}. For this, some changes are made in the original configuration presented in Section \ref{sec:background}, while maintaining 140 GHz as the frequency of operation. The changes come mainly in the size of the chiplets, reduced in this case to fit 16 in the same system dimensions, and in the location of the antennas. We distribute one antenna per chiplet. Similar to the analysis done in Section \ref{sub1sec5}, 
here we evaluate the ratio of the power in the target destination with respect to that of the rest of antennas, as well as the BER as a function of the data rate and the data rate at a specific BER. \par 
In Fig. \ref{fig:numchipletsa} it is observed that the CIR of the 4 chiplet case is longer compared to the 16 chiplet case. Therefore, as in Fig. \ref{fig:numchipletsb} the focused energy with TR is higher in 4 chips with a high energy concentration. Moreover, it is shown in Fig. \ref{fig:numchipletsb} that with TR 16 chiplet case could lead to comparatively high ISI with the less interference compressing surrounding the peak energy concentration.\par 

Fig. \ref{fig:numchipletsc} shows the comparison of the power ratio for 4 and 16 chiplets in two different links. For the four chiplet case we take the channel GC as example. To perform a fair comparison with the 16 chiplet case, we find a random link that are in a close physical position to the 4 chiplet counterparts, which would be links $HF$ in this case. In Fig. \ref{fig:cases}, we can see the position of antennas for the both cases.\par 
It is observed that for 16 chiplets the link HF concentrates poorly with TR. This result still bears an improvement concerning the non-TR version of it. However, compared to a channel in a similar position in the 4 chiplet version the concentration is not very good. This could be due to the distance, obstacles, and edges that are present in the 16 chiplet version, this could make the channel have more multipath components that are more difficult to compensate, therefore needing more power to achieve a successful transmission.\par 
The comparison made in Fig. \ref{fig:numchipletsd} for BER=0.1 further illustrates the difference between the two analyzed cases. This result indeed suggests that the more complex channel leads to the existence of higher correlations and, hence, higher interference even with TR. Also, the losses in the channel may be increasing, leading to a higher impact of the noise existing in all receivers. That would also explain the lower data rate in the single-link case.

\section{Performance Degradation with Non-Ideal Filters}

\label{sec:results3}

The spatial-temporal concentration of ideal TR filters has demonstrated superior performance on higher bandwidths up to 50 Gb/s thanks to the mitigation of ISI effects. However, so far we assumed that the TR filters have been implemented with perfect synchronized sampling of the pre-characterized CIR with infinite sampling speed. In practice, however, a digital filter implementing TR would need to have a finite sampling rate to be feasible.

Ideally, the filter needs to capture all the temporal variations of the channel to perfectly apply TR on a given channel. However, our full-wave simulations have shown that due to the high frequencies involved and the densely integrated environment being simulated, changes in the response in picosecond time intervals, which would need sampling frequencies in the THz range to be captured.
Implementation of filters with such resolution is unfeasible. 
On the other hand, quantization allows us to find a balance between the number of bits to be used to represent the amplitude levels of the TR filter, while maintaining the quality of the desired output. Therefore, based on realistic constraints, it is important to analyze the impact of TR with non-ideal filters and evaluate the spatial-temporal focusing based on the trade-off between the resource constraints and the received signal strength.

\subsection{Zero-Order Hold Filter} 

\begin{figure}[t!]
\centering
\includegraphics[width=0.7\columnwidth]{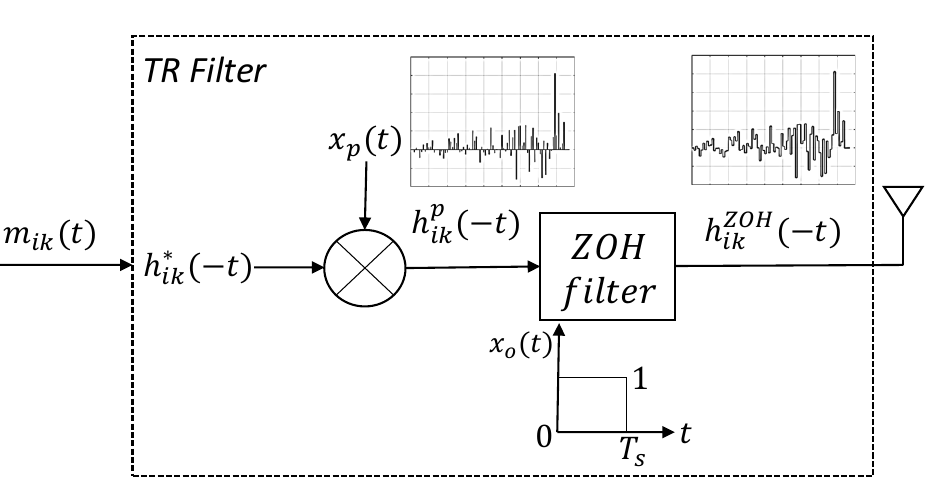}
\caption{Overview of zero-order hold filtering.}
\label{fig:filter}
\end{figure}

Inspired by the standard interpolation technique of sample and hold \cite{oppenheim}, here were propose a simple analysis considering the ideal TR estimation and hold the amplitude to the period of reduced sampling frequency. With this method, the sampled amplitudes are being held constant at the point of sampling operation is performed simply by introducing a synchronized square wave to the sampled signal with equal sampling frequency. This operation is identified as zero-order hold (ZOH) filtering. ZOH filter holds each sampled value for a duration of one sampling interval, as it is illustrated in Fig. \ref{fig:filter}.

Since the amplitude of the received multipath components is preserved until the next point of sampling while maintaining constant energy level on the particular sampling period, the ZOH filtering has minor degradation of the focused energy concentration with reduced sampling rates. We next evaluate the impact of reducing the sampling frequency.

\subsection{Performance Evaluation}
\begin{figure*}[t!]
\centering
\begin{subfigure}[t]{0.65\columnwidth}
\centering
\includegraphics[width=\textwidth]{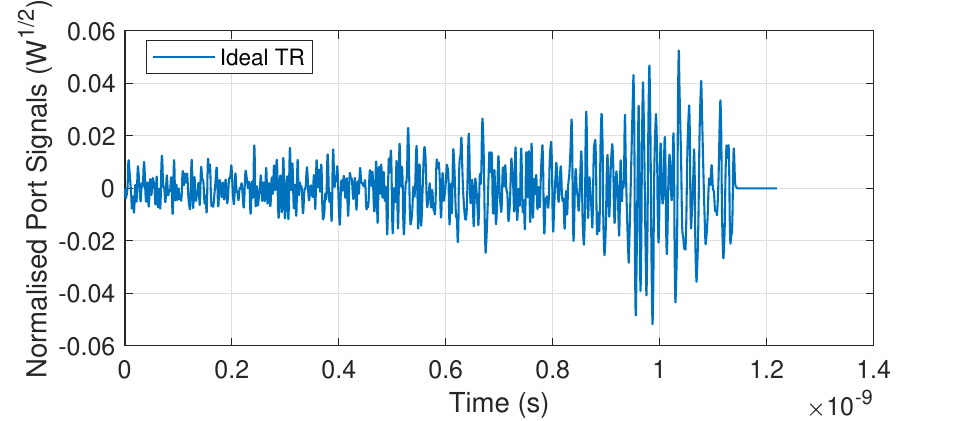}
\caption{}
\label{fig:scattera}
\end{subfigure}
\begin{subfigure}[t]{0.65\columnwidth}
\centering
\includegraphics[width=\textwidth]{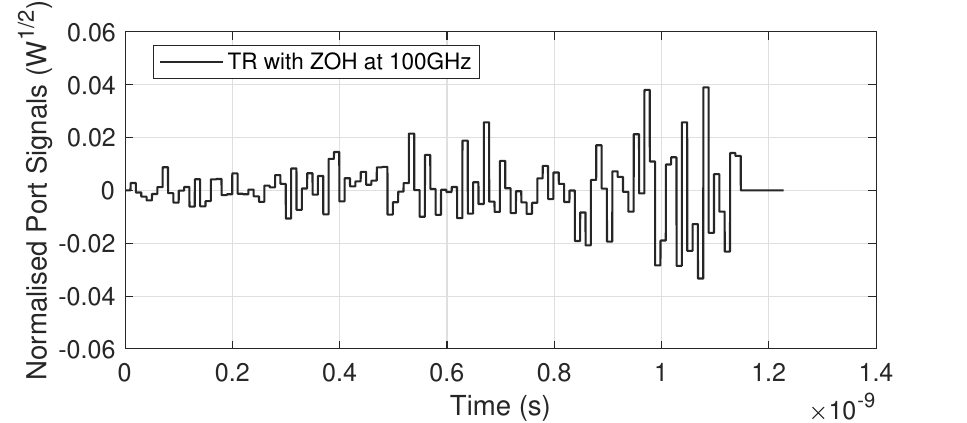}
\caption{}
\end{subfigure}
\begin{subfigure}[t]{0.65\columnwidth}
\centering
\includegraphics[width=\textwidth]{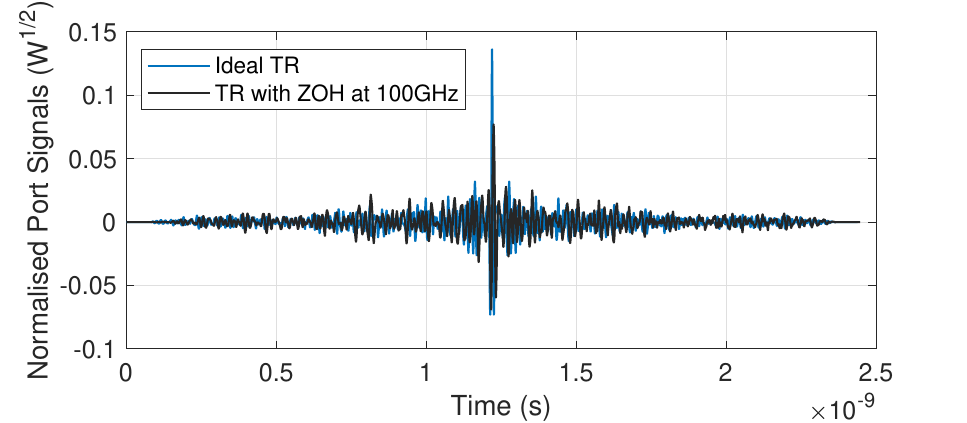}
\caption{}
\end{subfigure}
\vspace{-0.2cm}
\caption{Impact of the use of non-ideal filters on the response of the TR channel. (a) Response of the ideal TR filter. (b) Response of the TR filter with ZOH at 100 GHz sampling frequency. (c) Comparison of energy concentration with ideal TR filter and ZOH.}
\vspace{-0.4cm}
\label{fig:filterRes}
\end{figure*}

The general behavior of the ZOH filtering of the CIR is illustrated in Fig. \ref{fig:filterRes}. As can be observed, the ideal TR response which acts as the perfect filter has a quickly oscillating behavior. The second chart shows how the ZOH filter has reduced the sampling frequency down to 100 GHz, with noticeable non-linearities appearing due to the ZOH operation. Still, at 100 GHz, we still can see the high-energy components of the filter around 1 ns are still distinguishable. As ZOH acts as a typical low pass filter in the time domain, the frequency domain response corresponds to a $sinc(\cdot)$ function, which in return could limit the frequency components. Hence, if the sampling frequency is reduced further, significant portions of the CIR could be lost in the filter. Finally, Fig. \ref{fig:filterRes}(c) illustrates the difference between the convolution of the filter and the channel in both cases. We observe how the ideal TR has a higher energy in the peak and in the surroundings, whereas the ZOH version, even sampled at a very high frequency, has a slight broadening of the response.


\begin{figure*}[t]
\centering
\begin{subfigure}[t]{0.6\columnwidth}
\includegraphics[width=\textwidth]{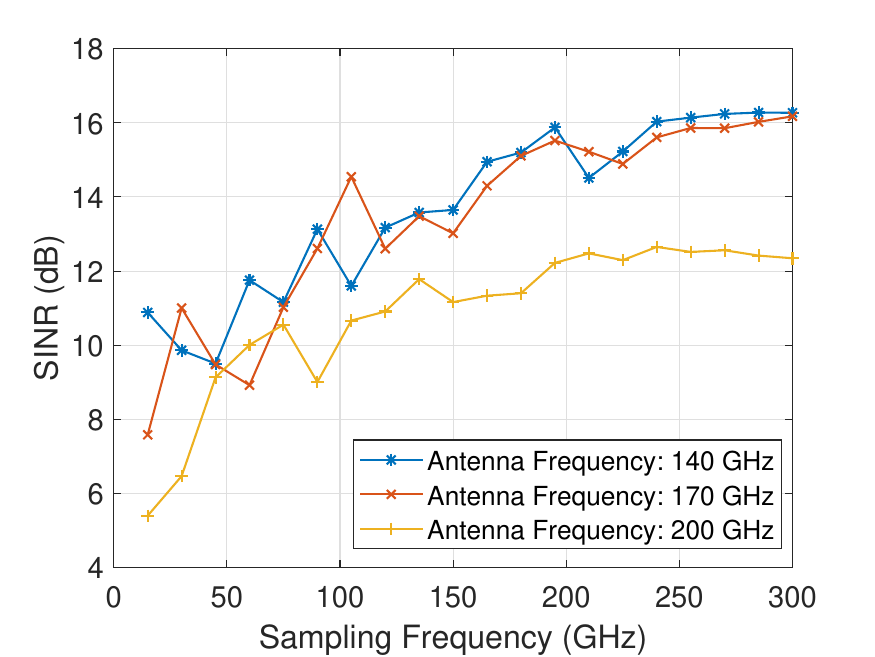}
\caption{}
\label{fig:scatterc1}
\end{subfigure}
\hspace{-0.2cm}
\begin{subfigure}[t]{0.6\columnwidth}
\includegraphics[width=\textwidth]{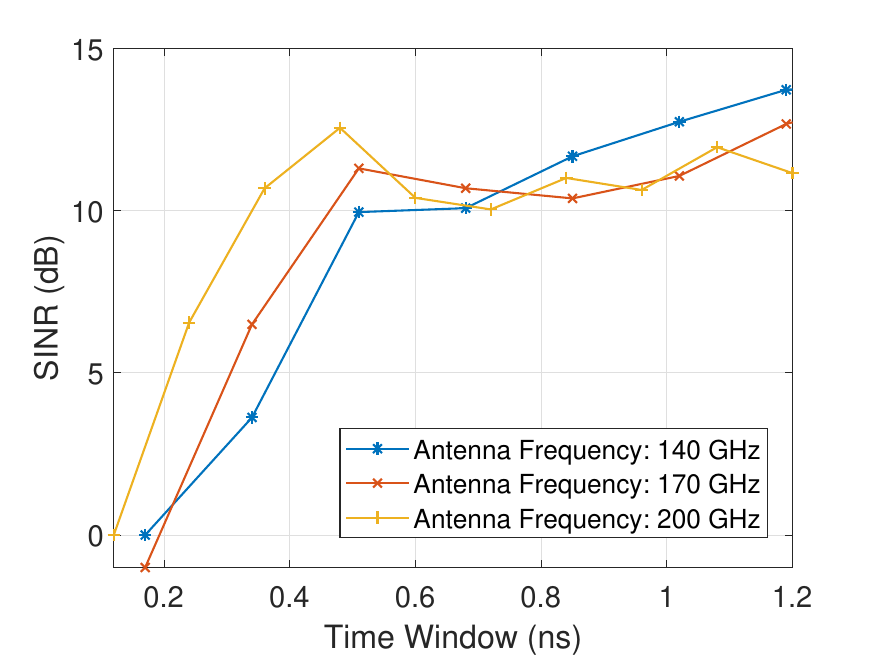}
\caption{}
\label{fig:scatterc2}
\end{subfigure}
\hspace{-0.2cm}
\begin{subfigure}[t]{0.6\columnwidth}
\includegraphics[width=\textwidth]{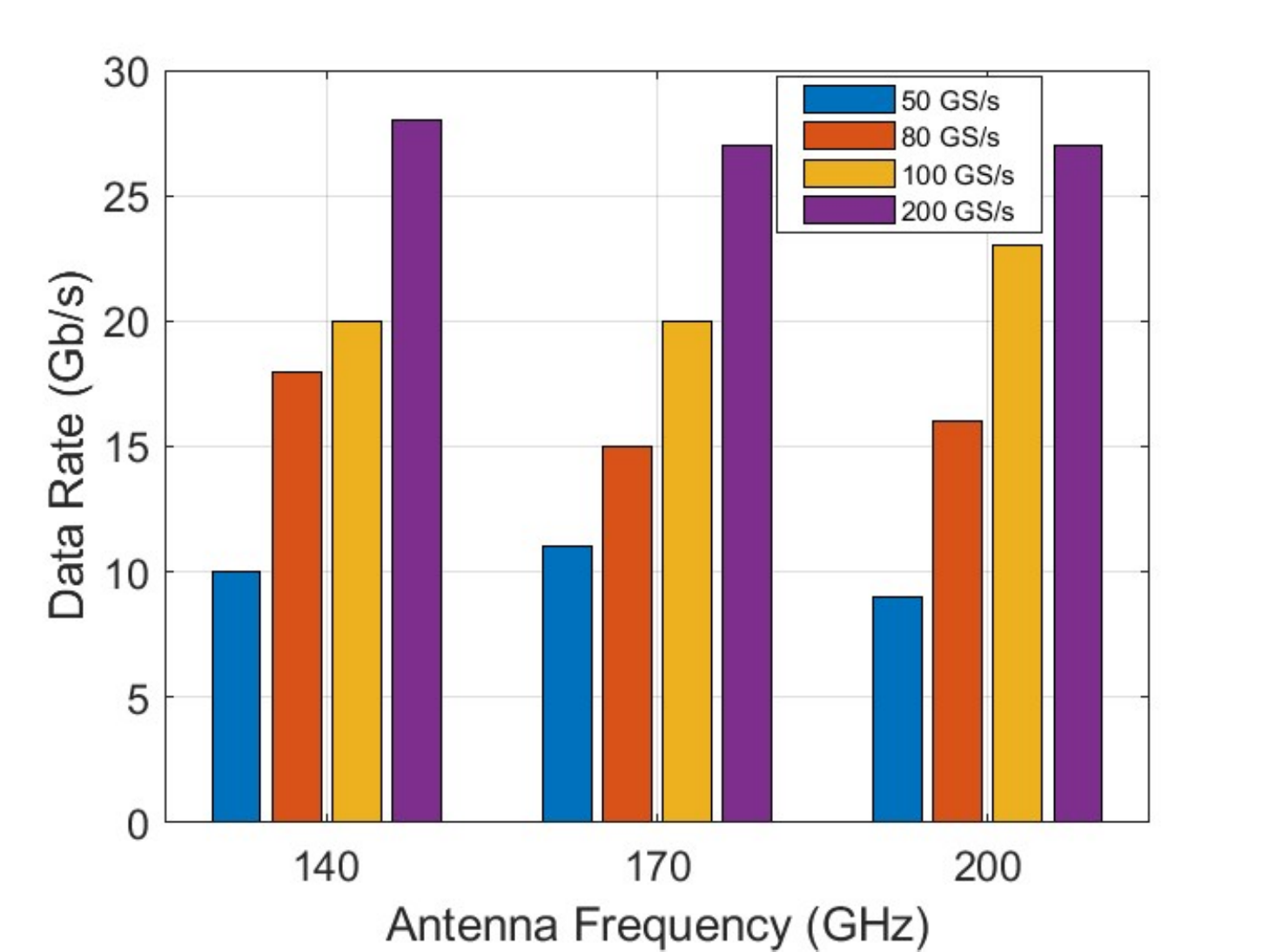}
\caption{}
\label{fig:scatterc3}
\end{subfigure}
\vspace{-0.2cm}
\caption{Impact of the TR filter properties on the performance of TR for the 140 GHz, 170 GHz, and 200 GHz antenna resonance frequencies. (a) $SINR_{stsr}$ within a window of 0.4 ns centered in the peak of the response, to that outside the window as a function of the sampling frequency.
(b) $SINR_{stsr}$ within a window of variable size to that outside the window as a function of the window size, for a sampling rate of 150 GHz.
(c) Achievable data rate as a function of the sampling rate of the TR filter at 0.1 BER and 10 dBm transmit power.}
\vspace{-0.4cm}
\label{fig:scatter2}
\end{figure*}

To evaluate the performance of the ZOH technique, we summarize the results of several evaluations in Fig. \ref{fig:scatter2}. First, we integrate the energy around the peak with a window of 0.4~ns (corresponding to a modulation speed of 2.5~Gb/s) as in Equation (\ref{eq12}) and compare it to the energy outside of the peak, as an indication of how well the temporal concentration is maintained when sampling at different frequencies. As shown in Fig. \ref{fig:scatter2}(a), the energy is rather constant until the sampling rate drops below 250~GHz. As we shift further down, there is a degradation of the ratio, yet without a drop of more than 6~dB until 50~GHz sampling frequency. 


Furthermore, the effect of interference due to mutipath was measured by varying the window size of the CIR for the interference measurement. This result provides us with an approximate idea on measured signal strength when the data rate is increased (and then the window reduced). From the results of Fig. \ref{fig:scatter2}(b), it is apparent that when the window size is increased, the amount of captured energy also increases proportionally. In the figure we see that the impact of the window size is more noticeable at small sizes, since the finite sampling rates causes the peak to broaden. The impact of the resonance frequency is mild.

Finally, to analyze the performance of the filter with BER, the data rates were evaluated with varied sampling frequencies (i.e. 200 GHz, 100 GHz, 80 GHz, and 50 GHz) for three resonance frequencies. As illustrated in Fig. \ref{fig:scatter2}(c) for a fixed BER of 0.1, the achievable data rates seem to degrade gracefully as we reduce the sampling rate by significant amounts, still achieving remarkable speeds for a sampling rate of 80 GHz. 

\section{Discussion}
\label{sec:Discussion}

TR filter is a novel approach for on-chip communication which simultaneously mitigate ISI and CCI, by leveraging the dispersive nature of the channel. Acting as a spatial matched filter, TR enhances the receive signal strength, while achieving  near-perfect energy focusing at the intended receiver. 
Despite the performance analysed from ideal to non-ideal filter implementation, it is important to note that in real hardware implementation could impose limitations and this work represents the performance metrics by considering the initial exploration of TR filter in diverse environments.

\subsubsection*{\textbf{Noise Figure at Transceiver}} 
\begin{figure*}[t]
\centering
\begin{subfigure}[t]{0.7\columnwidth}
\includegraphics[width=\textwidth]{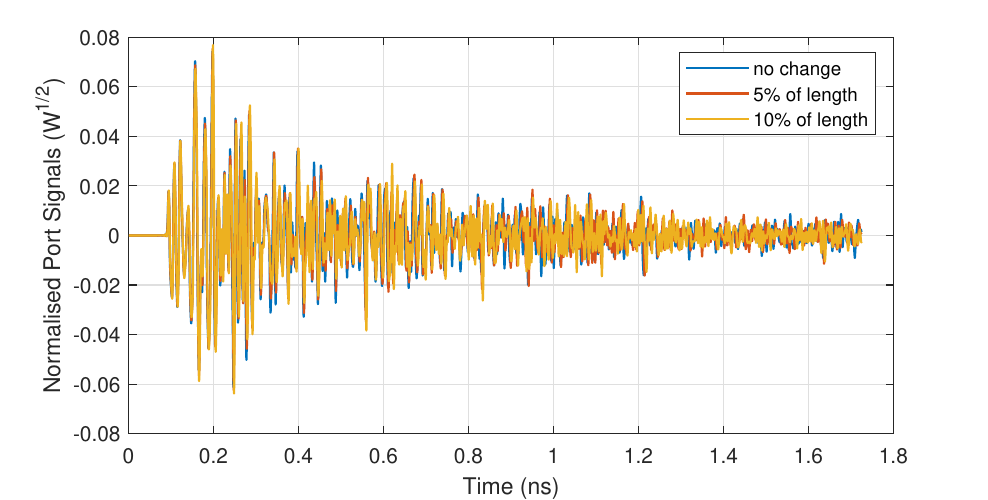}
\caption{}
\label{fig:var1}
\end{subfigure}
\hspace{-0.2cm}
\begin{subfigure}[t]{0.5\columnwidth}
\includegraphics[width=\textwidth]{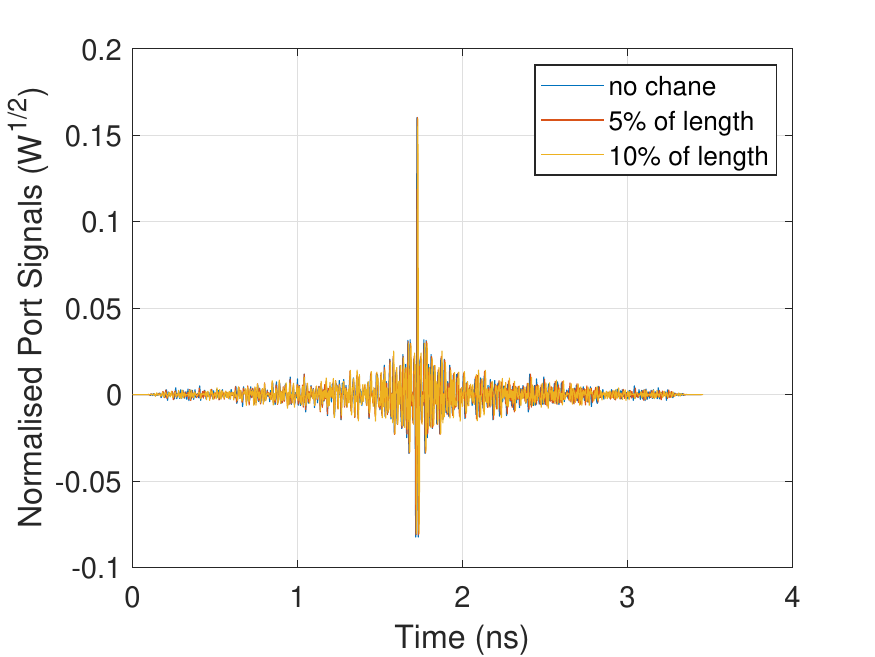}
\caption{}
\label{fig:var2}
\end{subfigure}
\hspace{-0.2cm}
\begin{subfigure}[t]{0.5\columnwidth}
\includegraphics[width=\textwidth]{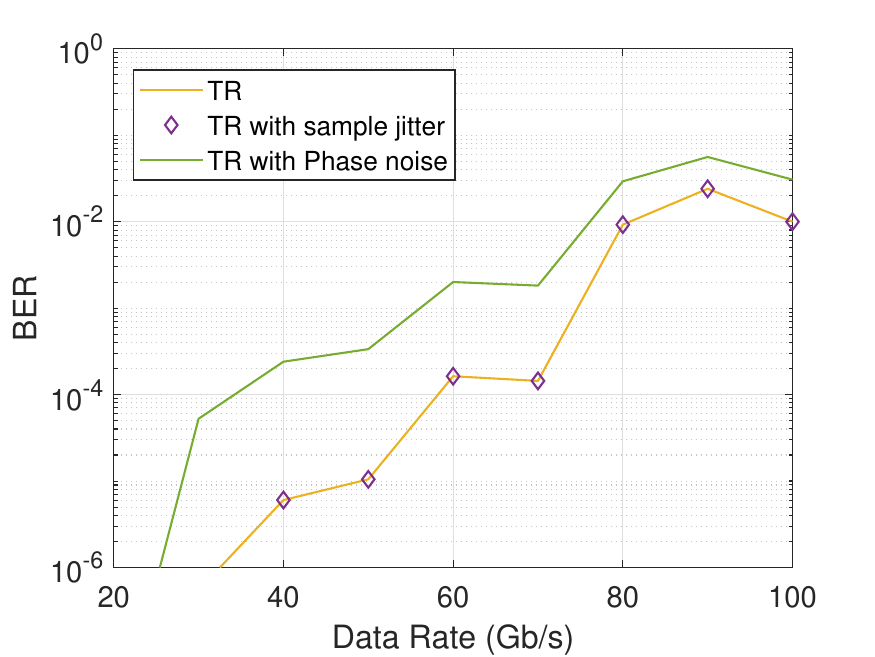}
\caption{}
\label{fig:phasenoise}
\end{subfigure}
\vspace{-0.2cm}
\caption{Baseline 140 GHz 4-chiplet interposer, (a) CIR with \% length expansion of Cu (b)  TR focused energy concentration with \% length expansion of Cu (c) Single link BER as a function of data rate with phase noise and sampling jitter (+0.5$T_{s}$) }
\vspace{-0.4cm}
\end{figure*}

The reported CMOS receiver front-ends in this frequency band \cite{Qahir2024, Wang2014, Shu2019} 
exhibit noise figures of 5 to 15 dB, phase noise around –110 to –80 dBc/Hz at 1 MHz offset, and DC–Effective Isotropic Radiated Power (EIRP) efficiencies of 10–20\%. Based on the data in \cite{Qahir2024}, we have modeled the phase noise as discussed in \cite{Nimr2023Radio} and the results are illustrated Fig. \ref{fig:phasenoise}. Moreover, by considering the sampling jitter, each sample of the CIR is moved +0.5 times the sampling period ($T_{s}$), to observe the effect of performance. The results in Fig \ref{fig:phasenoise} shows that the phase noise has more impact than the effect of jitter, as the sampling period is in nano-seconds in ideal TR filter response. However, these characteristics indicate that the real noise power should be scaled by roughly 10 dB with imperfect transceivers at 140 GHz (baseline configuration) and this work represents the first results representing the upper bound of TR filter performance.

\subsubsection*{\textbf{Area and Power}} In considering the area and power cost to implement the TR filter, the recent literature reports an on-chip 150 GS/s DAC \cite{moeneclaey2025} with an energy efficiency of 0.1656 aJ/bit. Moreover, as discussed in \cite{Ahmad2024}  and \cite{Guansheng2024}, DACs operating beyond 100 GS/s typically occupy an area in the range of 2.5-3.5 mm2. This capability provides us an open door to explore the sampling rates beyond 100 GS/s for TR filters, which gives data rates up to 25 Gb/s as discussed in  Section \ref{sec:results3}.

\subsubsection*{\textbf{Scalability}}
In this work, we analysed two chiplet configurations, (i) a 4 chiplet interposer with three antennas per each chiplet and  (ii) a 16 chiplet interposer with a single antenna per each chiplet. The 4-chiplet configuration enabled analysis of inter and intra-chiplet communication, including antenna coupling, interference management through collective spectrum sharing, and link performance. The 16-chiplet configuration was used to evaluate communication performance at higher chiplet densities, focusing on dense spatial focusing enabled by TR filtering.
Although these configurations differed in antenna density and chiplet count, the observed results exhibit qualitative and predictable trends from small to large scale systems \cite{shao2021}.

\subsubsection*{\textbf{Variability}}

Robustness is a crucial aspect in practical implementation of TR filters. The effective wavelength of the EM waves in a silicon interposer changes from 0.44-0.63 mm in the range of frequencies 140-200 GHz explored in this work. In considering geometry of our structure which is with $5 \times 5 \, mm^{2}$ chiplets, the dimensional and temperature variations, interconnect/bump randomness are in order of a few micro-meters, less than 0.5\% of the wavelength. To access the resilience, we have simulated a percentage expansion of length in Cu layer, considering it has the highest thermal expansion coefficient \cite{Touloukian1977}\cite{Hahn1970thermalexpansion} among the materials in a computing package. Considering the temperatures drift of a computer processor, the maximum expansion would be of 1-2\%. Yet to confirm the robustness, we simulate 5\% and 10\% of expansion. Fig. \ref{fig:var1} and Fig. \ref{fig:var2} show the measured CIR for the different lengths and the corresponding energy concentration with the original TR filter for the 140 GHz case.  It is observed that the figures show minor deviations based on the expansions with robust behavior upon variability to temperature drifts and dimension tolerance.


\section{Conclusion}

\label{sec:conclusion}

In this paper, we have presented the idea of applying TR filters to mitigate the sources of interference in static wireless communication scenarios within computing packages. To explore the effect of TR in varied setups, we evaluated the technique in a base scenario with four chiplets and antennas operating at 140 GHz, assuming ideal TR filters, as well as for multiple variations in frequency, geometry, and also employing TR filters with finite sampling rate.  
We observed that TR provides an order-of-magnitude improvement with respect to non-TR transmissions, and that allows to perform 2--3 concurrent transmissions to further boost the data rate. The results showed no clear frequency dependence because the environment was reverberant across frequency sweep, while increasing the number of chiplets caused significant changes due to the greater channel complexity with numerous changes of propagation medium. 
In future, we aim to explore the limits of the TR filter in terms of sampling and resolution, while analyzing the impact of hardware impairments on the transceiver. Moreover, we plan to, investigate system-level integration challenges, such as (i) achieving low-power \cite{Zhou2025} and low-area designs suitable for large-scale multi-chip deployments and (ii) how the existing TRMAC protocol \cite{bandara2025} can be co-optimized with the TR filter to improve efficiency and scalability in distributed systems.

\bibliographystyle{IEEEtran}
\bibliography{IEEEabrv,10_bib}

\section*{} 
\label{sec:bios}
\vspace{-1.5cm}
\begin{IEEEbiographynophoto}{Ama Bandara} 
is a PhD student at Universitat Polit\`{e}cnica de Catalunya (UPC), Barcelona, Spain. Her research interests include signal processing, wireless channel characterization, MAC protocol design and classical-quantum communications. She received her Master of Philosophy in Wireless Communications from The Open University of Sri Lanka, Colombo, Sri Lanka. She is a student member of IEEE Communications Society. Contact her at ama.peramuna@upc.edu.
\end{IEEEbiographynophoto}
\vspace{-1cm}
\begin{IEEEbiographynophoto}{F\'{a}tima Rodr\'{i}guez-Gal\'{a}n} 
is a PhD student at Universitat Polit\`{e}cnica de Catalunya (UPC), Barcelona, Spain. Her research interests include electromagnetics, antenna design, and wireless channel characterization. She received her Master in Advanced Communications Technologies Universidad Carlos III de Madrid, Spain. Contact her at fatima.yolanda.rodriguez@upc.edu.
\end{IEEEbiographynophoto}
\vspace{-1cm}
\begin{IEEEbiographynophoto}{Pau Talaran} 
is a Master student at Universitat Polit\`{e}cnica de Catalunya (UPC), Barcelona, Spain. His research interests include signal processing and filter techniques. He received his Bachelor in telecommunication engineering in Universitat Polit\`{e}cnica de Catalunya (UPC), Barcelona, Spain. Contact him at pau.talarn@estudiantat.upc.edu.
\end{IEEEbiographynophoto}
\vspace{-1cm}
\begin{IEEEbiographynophoto}{Elana Pereira de Santana} 
is a PhD student at the Institute of High Frequency and Quantum Electronics at University of Siegen. Her research interests include 2-D materials and graphene-based antennas for THz communications. She received her Master in Nanoscience and Nanotechnology from Universität Siegen, Siegen, Nordrhein-Westfalen, Germany. Contact her at Elana.PSantana@uni-siegen.de.
\end{IEEEbiographynophoto}
\vspace{-1cm}
\begin{IEEEbiographynophoto}{Evgenii Vinogradov} 
is a Director of Research at Universitat Politècnica de Catalunya in Spain. His research interests centered around wireless communication involving Non-Terrestrial Networks and the advancements of 5G-6G. He recieved his PhD in telecommunications engineering from UCLouvain in Belgium. Contact him at evgenii.vinogradov@upc.edu.
\end{IEEEbiographynophoto}
\vspace{-1cm}
\begin{IEEEbiographynophoto}{Peter Haring Bol\'{i}var} 
is the Chair for High Frequency and Quantum Electronics, University of Siegen. His research interests include terahertz technology, high-frequency electronics, nanotechnology, and photonics. He received his PhD in Semiconductor Electronics from RWTH Aachen University, Aachen, Germany. Contact him at peter.haring@uni-siegen.de.
\end{IEEEbiographynophoto}
\vspace{-1cm}
\begin{IEEEbiographynophoto}{Eduard Alarc\'{o}n} is a full professor at the Universitat Polit\`{e}cnica de Catalunya (UPC). His research interests include communications at the nano-scale, wireless power transfer and computer architecture. He received his PhD in Electrical Engineering from Universitat Polit\`{e}cnica de Catalunya (UPC), Barcelona, Spain. He is a senior member of the IEEE. Contact him at eduard.alarcon@upc.edu.
\end{IEEEbiographynophoto}
\vspace{-1cm}
\begin{IEEEbiographynophoto}{Sergi Abadal} 
is a Director and Distinguished Researcher at Universitat Polit\`{e}cnica de Catalunya (UPC), Spain. His research interests include graphene antennas, chip-scale wireless communications, and computer architecture. He received his PhD in Computer Architecture from the Department of Computer Architecture, Universitat Politècnica de Catalunya (UPC), Barcelona, Spain. He is a senior member of the IEEE and member of the ACM and HiPEAC. Contact him at abadal@ac.upc.edu.
\end{IEEEbiographynophoto}


\vfill

\end{document}